# Correlation transfer by layer 5 cortical neurons under recreated synaptic inputs *in vitro*

Daniele Linaro[1,2,3,4], Gabriel K. Ocker[5,6], Brent Doiron[6,7], Michele Giugliano[1,8]

[1] *Molecular, Cellular, and Network Excitability Lab, Department of Biomedical Sciences and Institute Born-Bunge, Universiteit Antwerpen, Wilrijk B-2610, Belgium*
[2] *Université Libre de Bruxelles (ULB), IRIBHM, and ULB Neuroscience Institute, Brussels, B-1050 Belgium*
[3] *VIB-KU Leuven Center for Brain & Disease Research, Leuven B-3000, Belgium*
[4] *Department of Neurosciences, Leuven Brain Institute, KU Leuven, Leuven B-3000, Belgium*
[5] *Department of Neuroscience, University of Pittsburgh, Pittsburgh, PA 15260, United States of America*
[6] *Center for the Neural Basis of Cognition, University of Pittsburgh and Carnegie Mellon University, Pittsburgh, PA 15213, USA*
[7] *Department of Mathematics, University of Pittsburgh, Pittsburgh, PA 15260, USA*
[8] *Scuola Internazionale Superiore di Studi Avanzati, Neuroscience sector, Trieste, Italy*

**Corresponding Author:** Dr. Michele Giugliano, michele.giugliano@uantwerpen.be
Universiteit Antwerpen
Campus Drie Eiken – T665
Universiteitsplein 1, 2610 Wilrijk (Belgium)

**Acknowledgments.** We are grateful to Drs. J. Couto, A. Bacci, V. Bonin, and K. Farrow for discussions, and to Mr. D. Van Dyck and M. Wijnants for excellent technical assistance. Financial support from the European Union's Horizon 2020 Framework Programme for Research and Innovation under the Specific Grant Agreement n. 785907 (Human Brain Project SGA2), the Belgian Science Policy Office (grant n. IUAP-VII/20), the Flemish Research Foundation (grants no. G0F1517N and no. K201619N), and Scuola Internazionale Superiore di Studi Avanzati ("Collaborazione di Eccellenza 2018") is kindly acknowledged, NIH grants 1U19NS107613-01, the Vannevar Bush faculty fellowship N00014-18-1-2002, and a grant from the Simons foundation collaboration on the global brain. The funders had no role in study design, data collection and analysis, decision to publish, or preparation of the manuscript.

**Abstract**

Correlated electrical activity in neurons is a prominent characteristic of cortical microcircuits. Despite a growing amount of evidence concerning both spike-count and subthreshold membrane potential pairwise correlations, little is known about how different types of cortical neurons convert correlated inputs into correlated outputs. We studied pyramidal neurons and two classes of GABAergic interneurons of layer 5 in neocortical brain slices obtained from rats of both sexes, and we stimulated them with biophysically realistic correlated inputs, generated using dynamic clamp. We found that the physiological differences between cell types manifested unique features in their capacity to transfer correlated inputs. We used linear response theory and computational modeling to gain clear insights into how cellular properties determine both the gain and timescale of correlation transfer, thus tying single-cell features with network interactions. Our results provide further ground for the functionally distinct roles played by various types of neuronal cells in the cortical microcircuit.

**Introduction**

Throughout the brain, the activity of neurons is correlated across multiple spatial and temporal scales (Cohen and Kohn, 2011, Doiron et al., 2016). In the cortex, neurons embedded in the same microcircuit show a high degree of functional similarity, not only in the correlations between spike trains (Erisken et al., 2014, Bair et al., 2001, deCharms and Merzenich, 1996), but also in sub-threshold voltage fluctuations (Poulet and Petersen, 2008, Gentet et al., 2010, Gentet et al., 2012). Several studies investigated the magnitude and timescales of spike-count correlations, known also as *noise correlations* (Cohen and Kohn, 2011). In primates, for example, correlations arise on relatively short timescales (i.e., up to a few hundred milliseconds) in the visual area MT (Bair et al., 2001) and their magnitude is strongly modulated both by the stimulus properties and internal brain state (Steinmetz et al., 2000).

The cortical origin of spike-count correlations has been investigated extensively *in vitro* and in theoretical studies. By artificially constraining the fraction of common inputs received by pairs of neurons, de la Rocha *et al.* (2007) demonstrated experimentally that spike-count correlations between spike trains increase with the mean firing rate of the cell pair. Albeit general, this result was based on current-clamp stimuli, which only roughly approximate the summed activity of a large number of excitatory and inhibitory inputs. The *in vitro* and *in silico* study of (Litwin-Kumar et al., 2011) removed this limitation, considering the high-conductance state of intact cortical circuits (Destexhe et al., 2001, Destexhe et al., 2003). The high-conductance state transferred short time scale (~10 ms) correlations better than the low-conductance state. However, the opposite was true for long time scale (~100 ms), prompting the overall effect of a shift in background conductance to be termed *correlation-shaping* (Litwin-Kumar et al., 2011). These theoretical results were further experimentally validated *in vitro*, although for only one value of the mean firing rate.

Theoretical arguments (de la Rocha et al., 2007, Doiron et al., 2016, Litwin-Kumar et al., 2011) explain these modulations in correlation transfer by associated shifts in the dynamical transfer functions of each cell in the pair (Kondgen et al., 2008, Linaro et al., 2018). This is consistent with studies showing that intrinsic properties—such as whether the neuron acts like an integrator or a coincidence detector (Konig et al., 1996)—also determine correlation transfer (Hong et al., 2012). It is well known that the input-output transfer and other intrinsic properties of neurons are quite distinct between cell classes (Connors and Gutnick, 1990, Markram et al., 2004). Nevertheless, despite this vast literature, little is known about the way in which different cell types transmit correlations.

Here, we present the results of dynamic-clamp experiments carried out in pyramidal cells and in fast-spiking (FS) and non-fast-spiking (NON-FS) interneurons from slices of rat cortical tissue. We found that, for low input correlation and across all cell types, the experimentally measured values of output covariance and their dependency on firing rate are in quantitative agreement with the values computed using linear response theory (Litwin-Kumar et al., 2011, Ocker and Doiron, 2014). Interestingly however, FS interneurons displayed significantly larger values of output covariance, making this cell type particularly suited for transmitting correlations to its postsynaptic targets. These results can be explained by considering single-cell properties—the steepness of the stationary rate-current curve, the membrane time constant and the degree of spike-frequency adaptation—and how they are modulated by dynamic-clamp stimuli. Indeed, by using single-compartment integrate-and-fire models that recapitulate the electrophysiological response properties of the three cell types considered here, we could produce covariance values that qualitatively replicate our experimental observations. Additionally, we found that in pyramidal cells the correlation-shaping mechanism introduced by (Litwin-Kumar et al., 2011) crucially depends on the properties of the stimulation paradigm.

In summary, our results highlight how the intrinsic properties of distinct neuronal types affect their correlated firing and hint at possible functionally distinct roles of different cell types in propagating correlations within cortical circuits.

## Materials and methods

### Brain tissue slice preparation

Experiments were performed in accordance with international and institutional guidelines on animal welfare. All procedures were approved by the Ethical Committee of the Department of Biomedical Sciences of the University of Antwerp (permission no. 2011_87), and licensed by the Belgian Animal, Plant and Food Directorate-General of the Federal Department of Public Health, Safety of the Food Chain and the Environment (license no. LA1100469). Wistar rats of either sex (2-4 weeks old) were anesthetized using Isoflurane (IsoFlo, Abbott, USA) and decapitated. Brains were rapidly extracted and immersed, to be sliced, in ice cold Artificial CerebroSpinal Fluid (ACSF), containing (in $mM$): 125 NaCl, 25 $NaHCO_3$, 2.5 KCl, 1.25 $NaH_2PO_4$, 2 $CaCl_2$, 1 $MgCl_2$, 25 glucose, saturated with 95% $O_2$ and 5% $CO_2$. Parasagittal sections (300 $\mu m$ thick) of the primary somatosensory cortex were cut using a vibratome (Leica VT1000 S, Leica Microsystems GmbH, Germany) and then incubated in ACSF at 36°C for at least 45 minutes. Slices were then stored at room temperature, until transfer to the recording chamber.

An upright microscope (Leica Microsystems, DMLFS), equipped with infrared Differential Interference Contrast videomicroscopy, was employed to visually identify layer 5 (L5) cortical neurons in a submerged slice recording chamber, under 60x magnification. Recordings were performed at $32 \pm 1$°C, under continuous perfusion with ACSF at a rate of 1 $mL/min$. All chemicals were obtained from Sigma-Aldrich (Diegem, Belgium).

**Electrophysiological recordings**

Patch-clamp recordings were obtained from the cell somata in the whole-cell configuration, employing glass pipettes pulled on a horizontal puller (P97, Sutter Instruments, Novato CA, USA) from filamented borosilicate glass capillaries (Hilgenberg, Malsfeld, Germany). Pipette electrode resistance was in the range $4 - 8\ M\Omega$, when filled with an intracellular solution containing (in $mM$): 115 K-gluconate, 20 KCl, 10 HEPES, 4 Mg-ATP, 0.3 $Na_2$-GTP, 10 $Na_2$-phosphocreatine, pH adjusted to 7.3 with KOH.

Recordings and current injections were performed with a single electrode by a current-clamp amplifier (EPC 10, HEKA Electronics, Lambrecht/Pfalz, Germany), but neither the bridge balance nor capacitance neutralization circuitries of the amplifiers were activated during the experiments. Instead, signal transfer properties of the glass microelectrodes were repeatedly estimated, by a linear non-parametric identification method (Brette et al., 2008), throughout each recording session. Quantified as the impulse response, the microelectrode transfer properties were employed to compensate for artifacts in the recorded membrane potential offline or online (i.e. in conductance-clamp experiments). All experiments were performed using the software toolbox LCG (Linaro et al., 2014). Liquid junction potentials were left uncorrected, but are not expected to affect our conclusions. Signals were sampled at a rate of $20\ kHz$ and digitized at 16 bits, with an A/D conversion board (NI PCI-6221, National Instruments, USA). The D/A converter of the same board was used to synthesize the external voltage-commands to the patch-clamp amplifier, at $20\ kHz$ and 16 bits of resolution.

**Current-clamp and conductance-clamp stimulation waveforms**

Our stimulation protocol closely follows the framework introduced in (de la Rocha et al., 2007), repeats it, and extends it to the case of conductance inputs, recreated by the dynamic-clamp technique (Chance et al., 2002, Robinson and Kawai, 1993) (Fig. 3A). While only one neuron was stimulated and recorded at a time, our entire set of experiments was examined *a posteriori* to study and quantify the similarity between the responses from pairs of (unconnected) neurons. In fact, synaptic inputs to these neuronal pairs had been synthesized *a priori* with a desired degree of similarity.

In the case of current-driven stimuli as in (de la Rocha et al., 2007), each of the currents $I_1(t)$ and $I_2(t)$ injected (non-simultaneously) into two neurons was defined as the sum of an independent term and of a common term:

(1)
$$I_1(t) = \mu_1 + \sigma_1 \cdot \left(\sqrt{1-c} \cdot \eta_1(t) + \sqrt{c} \cdot \eta_c(t)\right)$$
$$I_2(t) = \mu_2 + \sigma_2 \cdot \left(\sqrt{1-c} \cdot \eta_2(t) + \sqrt{c} \cdot \eta_c(t)\right)$$

where $c, \mu_1, \sigma_1, \mu_2, \sigma_2$ are numerical parameters varied experimentally, while $\eta_1(t), \eta_2(t)$, and $\eta_c(t)$ are randomly fluctuating waveforms, generated offline as independent realizations of an Ornstein-Uhlenbeck stochastic process (Uhlenbeck and Ornstein, 1930, Cox and Miller, 1965). These waveforms have Gaussian amplitude distribution, zero mean, unitary variance, and an auto-correlation function exponentially decaying with time constant of $\tau = 5\,ms$. They are generated as discrete-time approximations, by independently iterating for each sampling interval $\Delta t$ (i.e. $\Delta t = (20\,kHz)^{-1}$) the algebraic expressions $\eta_x(t + \Delta t) = (1 - \Delta t/\tau) \cdot \eta_x(t) + \sqrt{2\Delta t/\tau} \cdot \xi_x(t)$ with x ∈ {1,2, c}, where the quantities $\xi_1(t), \xi_2(t)$, and $\xi_c(t)$ are three independent sequences of Gaussian numbers obtained by employing three distinct initial seeds for a pseudo-random number generator (Press, 2007). By such a construction, $I_1(t)$ and $I_2(t)$ are realizations of colored Gaussian stochastic processes, with means $\mu_1$ and $\mu_2$, standard-deviations $\sigma_1$ and $\sigma_2$, respectively, and identical autocorrelation length $\tau = 5\,ms$. Importantly, their cross-correlation coefficient $c$, which is under the direct control of the experimenter, determines *a priori* the degree of similarity between the waveforms and represents, in a compact form, both the fraction of common inputs and the synchronous inputs coming from distinct presynaptic sources ($0 \leq c \leq 1$) (Fig. 3B).

In the case of conductance-driven stimuli, the total currents $I_1(t)$ and $I_2(t)$ were computer-generated in real time by conductance-clamp (Chance et al., 2002, Destexhe and Pare, 1999, Destexhe et al., 2001, Linaro et al., 2014) as a barrage of excitatory and inhibitory synaptic conductance inputs:

(2)
$$I_1(t) = G_{E,1}(t)(E_E - V_1(t)) + G_{I,1}(t)(E_I - V_1(t))$$
$$I_2(t) = G_{E,2}(t)(E_E - V_2(t)) + G_{I,2}(t)(E_I - V_2(t))$$

where $G_{E,1}(t), G_{E,2}(t)$ ($G_{I,1}(t), G_{I,2}(t)$) are randomly fluctuating excitatory (inhibitory) synthetic synaptic conductance waveforms, whose apparent reversal potential is chosen as $E_E = 0\,mV$ ($E_I = -80\,mV$) (Chance et al., 2002), and where $V_1(t)$ and $V_2(t)$ are the membrane potentials instantaneously recorded from the two cells, as shown schematically in Fig. 3A.

Similarly to Eq. 1, the excitatory (inhibitory) conductance waveform $G_{E,1}(t), G_{E,2}(t)$, ($G_{I,1}(t), G_{I,2}(t)$) were defined by the sum of an independent term and of a common term:

(3)
$$G_{E,1}(t) = \bar{g}_{E,1} + \sigma_{E,2} \cdot \left(\sqrt{1-c} \cdot \eta_{E,1}(t) + \sqrt{c} \cdot \eta_{E,c}(t)\right)$$
$$G_{E,2}(t) = \bar{g}_{E,2} + \sigma_{E,2} \cdot \left(\sqrt{1-c} \cdot \eta_{E,2}(t) + \sqrt{c} \cdot \eta_{E,c}(t)\right)$$
$$G_{I,1}(t) = \bar{g}_{I,1} + \sigma_{I,1} \cdot \left(\sqrt{1-c} \cdot \eta_{I,1}(t) + \sqrt{c} \cdot \eta_{I,c}(t)\right)$$
$$G_{I,2}(t) = \bar{g}_{I,2} + \sigma_{I,2} \cdot \left(\sqrt{1-c} \cdot \eta_{I,2}(t) + \sqrt{c} \cdot \eta_{I,c}(t)\right)$$

where $\eta_{E,1}(t), \eta_{E,2}(t), \eta_{E,c}(t), \eta_{I,1}(t), \eta_{I,2}(t), \eta_{I,c}(t)$, are six independent realizations of an Ornstein-Uhlenbeck stochastic process, generated offline as already described. These processes have heterogeneous auto-correlation time lengths ($\tau_E = 5\,ms$ and $\tau_I = 10\,ms$ for excitatory and inhibitory inputs, respectively), capturing the distinct decay kinetics of synaptic currents mediated by AMPA- and by GABA-A-receptors (Tuckwell, 1989). $\bar{g}_{E,1}$ and $\bar{g}_{E,2}$ ($\bar{g}_{I,1}$ and $\bar{g}_{I,2}$) and $\sigma_{E,1}$ and $\sigma_{E,2}$ ($\sigma_{I,1}$ and $\sigma_{I,2}$) represent the means and standard deviations of the total excitatory (inhibitory) synaptic conductances, for each of the cells in the pair. Under the diffusion approximation (Tuckwell, 1989), Eq. 3 mimics the collective effect of the asynchronous activation of a large number of presynaptic excitatory (inhibitory) afferents, each with a corresponding a unitary excitatory (inhibitory) postsynaptic peak conductance $g_{EPSC}$ ($g_{IPSC}$). If the presynaptic mean frequency of excitatory (inhibitory) activation is indicated by $R_E$ ($R_I$), then $\bar{g}_E$ ($\bar{g}_I$) and $\sigma_E$ ($\sigma_I$) can be expressed as (Destexhe and Bal, 2009):

$$
\begin{aligned}
\overline{g}_E &= g_{EPSC} \cdot \tau_E \cdot R_E & \sigma_E &= \sqrt{0.5 \cdot g_{EPSC}^2 \cdot \tau_E \cdot R_E} \\
\overline{g}_I &= g_{IPSC} \cdot \tau_I \cdot R_I & \sigma_I &= \sqrt{0.5 \cdot g_{IPSC}^2 \cdot \tau_I \cdot R_I}
\end{aligned}
$$
(4)

In all our experiments, $g_{EPSC}$ was fixed to 2% of the inverse of the input resistance of the cell, while $g_{IPSC}$ was fixed to 6% of the same value, as proposed in (Chance et al., 2002). The mean firing rate of the presynaptic inhibitory population was fixed to $R_I = 3000\ Hz$, while $R_E$ was determined as the value balancing the mean excitatory and inhibitory drives, whenever the postsynaptic membrane potential fluctuates around a value $V_0$. By construction, at such a value $V_0$ of the membrane potential, the mean current injected into each neuron (Eq. 2) vanishes:

$$
0 = \langle I(t) \rangle \approx \overline{g}_E \cdot (E_E - V_0) + \overline{g}_I \cdot (E_I - V_0)
$$
(5)
$$
\Leftrightarrow \quad R_E = [g_{IPSC} \cdot \tau_I \cdot R_I \cdot (E_I - V_0)] / [g_{EPSC} \cdot \tau_E \cdot (E_E - V_0)]
$$

By this definition, acting on the value of a single parameter $V_0$ allows one to change the ratio between excitatory and inhibitory inputs and thus to alter the output firing rate of the patched neuron. In analogy to the current-driven synaptic inputs (Eq. 1), where $\mu, \sigma$ were employed to increase or decrease the neuronal firing rate of each patched cell, $V_0$ was employed to study the dependence of the correlation transfer on the neuronal firing rate in the case of conductance injection.

All electrophysiological experiments were carried out with LCG (Linaro et al., 2014).

**Experimental design**

As in (de la Rocha et al., 2007), for each recorded neuron we delivered *blocks* of $N = 100$ stimulation trials, each lasting 1.1 $s$ and interleaved by inter-stimulus intervals lasting 1-6 $s$, depending on the elicited firing rate (i.e. the higher the rate the longer the resting interval). Each block was repeated for 1–5 times, during which the parameters of the stimulation (e.g. $c$ and the pair $\mu, \sigma$ or the value of $V_0$) were kept constant. This series of repetitions was termed a *stimulus set*, and each neuron was presented with 1-7 distinct stimulus sets, differing in the values chosen for $c$ and $\mu$, $\sigma$, or for $c$ and $V_0$.

The identification of each patch electrode transfer properties, required for the Active Electrode Compensation, was performed once at the beginning of each block (i.e. every 5-10 min of recording), although in some instances it was repeated halfway through the block.

The initial seeds of the pseudo-random number generator, used for generating the shared component of each stimulus, were identical across the repetitions of the stimulation blocks and across cells. Within a block, each trial was generated by a distinct seed, so that a total of $N$ seeds for generating $\eta_c(t)$ in Eq. 1 and of $2N$ seeds for generating $\eta_{E,c}(t)$, $\eta_{I,c}(t)$ in Eq. 3 were chosen once for all the experiments. On the other hand, the seeds of all the independent components (i.e. $\eta_1(t), \eta_2(t)$ in Eq. 1; $\eta_{E,1}(t), \eta_{E,2}(t), \eta_{I,1}(t), \eta_{I,2}(t)$ in Eq. 3) were never fixed, but instead varied across trials, repetitions, and cells. All LCG scripts, configuration files, and command line strings required to precisely replicate our experimental protocols, together with the MATLAB scripts for the offline analysis of the recorded data, are available from FigShare.com.

After establishing the whole-cell configuration, in a subset of the experiments (Figures 1 and 5) we initially also probed the steady-state frequency-current curve of the neuron as in (Chance et al., 2002). Specifically, we injected $1\,s$-long DC depolarizing current steps increasing amplitude, superimposed to conductance-driven recreated synaptic inputs, and measured the evoked firing rate as the number of emitted spikes divided by $0.9\,s$, after discarding the first $0.1\,s$ of the response to each stimulation step. Equation 3 was used to generate the synaptic inputs, with $c$ set to zero and $\eta_{E,1}(t), \eta_{E,2}(t), \eta_{I,1}(t), \eta_{I,2}(t)$ generated as independent realizations. The value of $V_0$ was fixed so that the neuron fired on the average at ~$0.5\,Hz$, in the absence of any additional DC current step. This ensured that the modulation effect of the conductance-driven inputs was purely divisive for the frequency-current curve (Chance et al., 2002).

**Data analysis: passive and active membrane electrical properties**

Data were analyzed off line, using custom MATLAB scripts (The Mathworks, Natick MA, USA). Passive membrane electrical properties, such as the input resistance and membrane time constant, were evaluated. The time constant was estimated as the longest time constant of a double exponential function, best fitted to the repolarizing phase following the application of a $10\,ms$-long pulse of $-300\,pA$ amplitude. The input resistance was estimated as the slope of a steady-state current-voltage relationship, obtained by injecting $3\,s$-long subthreshold step current pulses with increasing amplitude. The rheobase current amplitude was estimated as the minimal DC stimulus amplitude necessary to elicit sustained firing.

The "sag" in the transient membrane potential trajectory in response to the injection of a $3\,s$-long hyperpolarizing step of current, was quantified as the ratio $\Delta V_{ss}/\Delta V_{peak}$: the steady state voltage deflection $\Delta V_{ss}$ was computed as the difference between $V_{rest}$, the resting potential averaged over the $500\,ms$ preceding the step onset, and the (hyperpolarized) potential averaged in the final $500\,ms$ of the step; $\Delta V_{peak}$ was computed as the difference between the above defined $V_{rest}$ and the minimum hyperpolarized potential, within the $300\,ms$ following the onset of the step. The amplitude of the hyperpolarizing current was chosen such that $\Delta V_{ss}$ was $\sim 10\,mV$.

The times of occurrence $\{t_q\}$ of the spikes fired by the neuron, in response to a given stimulation trial, were detected by finding the peaks of the intracellular voltage traces exceeding a threshold of $-20\,mV$. Other quantities were also extracted from each spike, such as (i) its threshold, defined as the voltage corresponding to the maximum of the third derivative of the membrane potential (Gentet et al., 2010, Henze and Buzsaki, 2001, Kole and Stuart, 2008); (ii) its amplitude, defined as the difference between its peak amplitude and the threshold; and (iii) its duration, measured as the width of the spike waveform at its half-amplitude.

In order to quantify the degree of spike frequency adaptation across cortical cell types, the accommodation index $A$ was also computed as (Druckmann et al., 2007, Shinomoto et al., 2003)

$$(6) \quad A = (N_{spks} - 4 - 1)^{-1} \cdot \sum_{q=4}^{N_{spks}-1} (\text{ISI}_q - \text{ISI}_{q-1})/(\text{ISI}_q + \text{ISI}_{q-1})$$

where $N_{spks}$ is the number of spikes evoked by a constant depolarizing $1\,s$-long step of current, $\text{ISI}_q$ is the $q$-th inter-spike interval (i.e. $\text{ISI}_q = t_{q+1} - t_q$), and where the first four spikes were always discarded from the analysis (i.e. the sum starts from $q = 4$). The amplitude of the step of current was chosen such that it elicited 15, 50 and 20 spikes/s for pyramidal, FS and NON-FS cells, respectively.

**Experimental estimation of the spike-count covariance, in neuronal pairs**

For estimating the covariance of the spike counts, we closely followed the procedure outlined in (de la Rocha et al., 2007). In detail, under current-driven and conductance-driven recreated synaptic inputs, and after discarding the first $0.1\ s$ of each stimulation trial to minimize transient effects, we detected the spike times $\{t_1, t_2, \ldots, t_q, \ldots\}$ over the remaining interval of duration $L = 1\ s$. For each trial, the spike times were equivalently expressed as a discrete-time spike train, converting them into a binary string $y(j)$, $j = [0, 1, 2, \ldots, \lfloor \frac{L}{\Delta t} \rfloor]$ upon uniform discretization of time, with a step of $\Delta t = 1.2\ ms$, where $\lfloor x \rfloor$ indicates the largest integer not greater than $x$:

$$(7) \quad y(j) = \begin{cases} 1 & \text{if there exists (a spike) } q: t_q \in [\,j \cdot \Delta t\,;(j+1) \cdot \Delta t\,) \\ 0 & \text{otherwise} \end{cases}$$

The discrete-time spike train $y(j)$ was then used to estimate the spike count, computed over a sliding window of duration $T$ (Dayan and Abbott, 2001):

$$(8) \quad n(j) = \sum_{i=j}^{j+\lfloor T/\Delta t \rfloor} y(i) \quad \text{with} \quad j = \{0, 1, 2, \ldots, M\}$$

where $M = \lfloor (L - T)/\Delta t \rfloor$.

The spike count was obtained for every trial in a block of $N$ stimulations, for every block out of $R$ repetitions, and for every cell. In the following, we use the notation $n_1^{k,r}(j)$ and $n_2^{k,r'}(j)$ to indicate, for two neurons, the spike counts of the response of cell 1 and cell 2, computed for the $k$-th trial and the $r$-th (or $r'$-th) repetition (i.e. as the number of repetitions $R_1$ and $R_2$ may be different across cells). As in (de la Rocha et al., 2007), the estimate of the covariance of the spike counts of the cell pair, obtained during the corresponding trial, was corrected for the time-shift and averaged across all $R_1$ and $R_2$ repetitions and $N$ trials[1] as:

(9)
$$\text{Cov}(n_1, n_2) = \frac{1}{N \cdot R_1 \cdot R_2} \sum_{k,r,r'}^{N,R_1,R_2} \frac{\left(\sum_{j=0}^{M} n_1^{k,r}(j) \cdot n_2^{k,r'}(j)\right)}{(M+1)} - \frac{\left(\sum_{j=0}^{M} n_1^{k,r}(j) \cdot n_2^{k+1,r'}(j)\right)}{(M+1)},$$

Finally, the correlation coefficient $\rho_T$ between the response of two neurons was obtained from the spike-count covariance as:

(10)
$$\rho_T = \frac{\text{Cov}(n_1, n_2)}{\sqrt{\text{Var}(n_1)\text{Var}(n_2)}},$$

with $\text{Var}(n_i) = \text{Cov}(n_i, n_i)$. Throughout our work, we used the spike-count covariance, except for section "Output correlation depends on the fraction of common inputs" (Fig. 3), where the linear relationship between input and output correlations is discussed.

Finally, as already mentioned in the previous sections, each cell was presented with several *stimulus sets*, each characterized by distinct values for $c$ and $\mu, \sigma$ or $c$ and $V_0$. As those parameters affect the mean spiking rate of the neurons, care was required to choose comparable responses across cells, out of the entire experimental data set. Therefore, we computed the correlation coefficient only for those stimulus-set pairs $l, m$ that elicited firing rates $\nu_1$ and $\nu_2$ (estimated over the whole stimulus set) that did not differ by more than 50%, i.e., $0.5 \leq \nu_1^l / \nu_2^m \leq 2$. We also tested more stringent conditions on the ratio $\nu_1^l / \nu_2^m$ to select the stimulus-set pairs to include in the analysis (up to $0.8 \leq \nu_1^l / \nu_2^m \leq 1.25$) and found that this does not alter qualitatively the dependence of the covariance on the geometric mean firing rate of the cell pair.

---

[1] In order to make the notation consistent, it is intended that across the sum over $k$ there is one trial after the last, which equals the first, i.e. $n_2^{M+1,r'} = n_2^{1,r'}$.

## Statistical analysis

Unless otherwise noted, all data are presented as mean ± standard deviation and statistical significance was assessed using a two-sample Kolmogorov-Smirnov test (Press, 2007).

## Theoretical prediction of the covariance values

We compared the empirical spike-count covariance to the values obtained from a theoretical prediction. This prediction is based on a well-established linear response theory for spike train covariability (de la Rocha et al., 2007). We first indicate in continuous time the spike train $y(t)$ as a sum of Dirac's Delta functions,

$$(11) \qquad y(t) = \sum_q \delta(t - t_q)$$

and the corresponding spike count $n(t)$ as its convolution with the *window* function $w_T(t)$,

$$(12) \qquad n(t) = \int_t^{t+T} y(x)dx = y(t) * w_T(t),$$

where $w_T(t) = 1$, for $t \in [-T\,;0]$ and otherwise zero. The cross-covariance function of the spike counts of the two neurons receiving common input can be expressed in terms of the cross-covariance function of the spike trains (Cox and Isham, 1980) as

$$(13) \qquad \text{Cov}(n_1, n_2)(\tau) = \text{Cov}(y_1, y_2)(\tau) * \Delta_T(\tau)$$

where $\Delta_T(\tau)$ is an even function defined as $\Delta_T(\tau) = \int_{-\infty}^{+\infty} w_T(s)w_T(\tau + s)ds$ taking the value of $T - |\tau|$, for $\tau \in [-T\,;T]$, and otherwise zero.

By the Wiener-Khinchin theorem, the cross-covariance function $\text{Cov}(y_1, y_2)(\tau)$ is the Fourier anti-transform of the cross-spectrum of the spike trains $\widehat{\text{Cov}}(\hat{Y}_1, \hat{Y}_2)(\omega)$::

$$(14) \qquad \text{Cov}(y_1, y_2)(\tau) = \frac{1}{2\pi} \cdot \int_{-\infty}^{+\infty} \widehat{\text{Cov}}(\hat{Y}_1, \hat{Y}_2)(\omega)\, e^{j\omega\tau} d\omega$$

Considering current-clamp stimuli as in de la Rocha *et al.* (2007), and assuming a small amplitude for the common input $q(t)$ (i.e. $\sigma \cdot \sqrt{c}$), we approximate the instantaneous spiking responses of the neurons as dominated by linear dynamical transfer properties. In the Fourier domain this corresponds to (Brunel et al., 2001)

$$\hat{Y}_1(\omega) \cong \hat{Y}_{1,0}(\omega) + \hat{A}_1(\omega) \cdot \hat{Q}(\omega)$$
$$\hat{Y}_2(\omega) \cong \hat{Y}_{2,0}(\omega) + \hat{A}_2(\omega) \cdot \hat{Q}(\omega)$$

(15)

where $\hat{Q}(\omega)$ is the Fourier transform of $q(t)$, $\hat{Y}_{i,0}(\omega)$ corresponds to the baseline spike train (when $q(t) = 0$, i=1,2), and where $\hat{A}_1(\omega)$ and $\hat{A}_2(\omega)$ are the dynamical response functions of the two neurons. For these quantities, a full experimental characterization was demonstrated previously (Higgs and Spain, 2009, Ilin et al., 2013, Kondgen et al., 2008, Linaro et al., 2018).

Assuming that the two neurons' spike trains are conditionally independent, given the common input $q(t)$, we evaluate (Holden, 1976) and approximate the cross-spectrum by Eq. 15 and obtain

(16)
$$\widehat{Cov}(\hat{Y}_1, \hat{Y}_1)(\omega) = \langle \hat{Y}_1^*(\omega) \cdot \hat{Y}_2(\omega) \rangle \cong \hat{A}^*_1(\omega) \cdot \hat{A}_2(\omega) \cdot \hat{S}(\omega)$$

where $\hat{A}^*_1(\omega)$ is the complex conjugate of $\hat{A}_1(\omega)$, and $\hat{S}(\omega) = \langle \hat{Q}^*(\omega) \cdot \hat{Q}(\omega) \rangle$ is the power spectrum of the common input $q(t)$.

Finally, estimating the covariance of the spike counts $n_1(t)$ and $n_2(t)$, requires evaluating Eq. 13 in $\tau = 0$ and substituting eq. 16 into eq. 14:

(17)
$$Cov(n_1, n_2)(0) \cong \frac{1}{2\pi} \cdot \int_{-\infty}^{+\infty} \hat{A}^*_1(\omega) \cdot \hat{A}_2(\omega) \cdot \hat{S}(\omega) \cdot \widehat{\Delta_T}(\omega) d\omega$$

where $\widehat{\Delta_T}(\omega)$ is the Fourier transform of $\Delta_T(\tau)$, and takes the form of $\widehat{\Delta_T}(\omega) = T^2 \sin^2(\omega T/2)/(\omega T/2)^2$. For large values of $T$, $\widehat{\Delta_T}(\omega)$ approaches a Dirac's Delta function $\widehat{\Delta_T}(\omega) \approx 2\pi T \cdot \delta(\omega)$, thereby canceling the contribution to the integral for values of the integrand far from $\omega \approx 0$. In addition, the dynamical response of cortical cells is rather constant at low values of the Fourier frequencies (Fourcaud and Brunel, 2002, Kondgen et al., 2008, Testa-Silva et al., 2014, Linaro et al., 2018). For these two reasons, $\hat{A}_1(\omega)$ and $\hat{A}_2(\omega)$ are further approximated by their value in $\omega \approx 0$, which takes the form of the local slope (i.e., the steady-state gain) of the stationary rate-current curve $F(I)$, computed at the cell's mean output firing rate $\nu$ (Brunel et al., 2001, Chance et al., 2002).

$$\text{(18)} \quad \text{Cov}(n_1, n_2)(0) \approx \left.\frac{d\,F_1(I)}{dI}\right|_{\nu_1} \cdot \left.\frac{d\,F_2(I)}{dI}\right|_{\nu_2} \cdot \frac{1}{2\pi} \cdot \int_{-\infty}^{+\infty} \hat{S}(\omega) \cdot \widehat{\Delta_T}(\omega) d\omega$$

The covariance is therefore expected to be proportional to the product of the slopes of the rate-current curves of the two cells. For some experiments, we thus probed the rate-current curve of each neuron, in addition to the other stimulation protocol, and we employed its local slope to approximate the neuron linear response function at the low frequencies needed to estimate spike-count covariances over long windows. In fact, although fast experimental techniques for the characterization of $\hat{A}_1(\omega)$ and $\hat{A}_2(\omega)$ are available (Higgs and Spain, 2009, Ilin et al., 2013), using the full transfer function is anyway not practical: $\hat{A}_1(\omega)$ and $\hat{A}_2(\omega)$ are in fact modulated by the mean firing rate $\nu_1$ and $\nu_2$ of the neurons (Linaro et al., 2018) and their complete estimate, across a sufficiently wide range of firing rates, is incompatible with the limited duration of each experiment.

**Neuron models**

For modeling different cortical cell types, we used an adaptive exponential integrate-and-fire model neuron (Brette and Gerstner, 2005), modified to incorporate a slow voltage-gated adaptation current $I_{slow}$. The neuron subthreshold membrane potential V evolves in time as

$$\text{(19)} \quad C\frac{dV}{dt} = \left[\bar{g} \cdot (V_L - V) + \bar{g} \cdot \Delta \cdot e^{(V-V_T)/\Delta}\right] + I_{slow} + I_{ext}$$

$$I_{slow} = \bar{g}_w \cdot w \cdot (V_w - V) \quad \tau_w \frac{dw}{dt} = \left(1 + e^{-(V-V_{wh})/D_w}\right)^{-1}$$

$$I_{ext} = G_E(t)(E_E - V) + G_I(t)(E_I - V)$$

where $C$ and $\bar{g}$ are the capacitance and conductance per unit of membrane surface, respectively, $V_L$ the resting potential, $\Delta$ the spike steepness and $V_T$ the spike initiation threshold (Fourcaud and Brunel, 2002). $g_w$ represents the peak adaptation conductance, $w$ its activation state variable and $V_w$ its reversal potential. The steady-state activation of the adaptation conductance is a logistic function, monotonically increasing with the membrane potential, with half-maximal activation at $V_{wh}$ and activation slope $D_w$. As soon as V crosses the peak threshold $V_{th}$, it repolarizes linearly to its resting value $V_L$, over a (refractory) period $\tau_{ref}$.

In order to capture the features of each cortical cell type, we adapted the model parameter to reproduce the resting membrane potential, the accommodation index, the rheobase, and the slope of the rate-current relationship measured experimentally. We specifically adjusted the values of $V_L$, $C$ and $\bar{g}$ to match the resting membrane potential and membrane time constant for each neuron type. The spike initiation threshold of the model, $V_T$, was adjusted to ensure that the models had the same ordering of rheobases as in the experiments (i.e., NON-FS < Pyr < FS). While the resulting ordering of $V_T$ was NON-FS < FS < Pyr, the value $\bar{g}_w$ of the adaptation conductance and the leak reversal potential also affect the rheobase.

The peak adaptation conductance, $g_w$, was then chosen to approximately match the accommodation indices for each neuron type, upon injecting a depolarizing constant current as in the experiments. All the parameters were identical across cell types for the three distinct model cells, as summarized in Table 1.

**Table 1:** Numerical parameters employed in the integrate-and-fire model neurons.

| Parameter | Description | Pyramidal | FS | NON-FS |
|---|---|---|---|---|
| $C$ | Membrane capacitance | 1 uF/cm2 | 1 | 1 |
| $\tau$ | Passive membrane time constant | 25 ms | 15 ms | 25 ms |
| $V_L$ | Leak reversal potential | -65 mV | -80 mV | -65 mV |
| $g_w$ | Peak Kv7 conductance | 0.002 mS/cm² | 0 mS/cm² | 0.1 mS/cm² |
| $\tau_w$ | Kv7 activation time constant | 200 ms | 200 ms | 200 ms |
| $V_{wh}$ | Kv7 half-activation voltage | -40 mV | -40 mV | -40 mV |
| $D_w$ | Slope of Kv7 activation | 8 mV | 8 mV | 8 mV |
| $V_w$ | Kv7 reversal potential | -85 mV | -85 mV | -85 mV |
| $\Delta$ | AP steepness | 1.4 mV | 0.2 mV | 1.4 mV |
| $V_T$ | AP initiation threshold | -46 mV | -50 mV | -52 mV |
| $V_{th}$ | AP threshold | 0 mV | 0 mV | 0 mV |

| $\tau_{ref}$ | Absolute refractory period | 6 ms | 1 ms | 4 ms |

With the aim of reproducing the conductance-clamp protocols, the external current density $I_{ext}$ was synthesized as in the experiments by Eq. 2. For simplicity however, we set $g_{EPSC} = g_{IPSC} = 0.06\,mS/cm^2$, and considered instantaneous synaptic coupling (i.e. $\tau_E, \tau_I \to 0$) (Tuckwell, 1989), and $R_I = 10\,kHz$, while changing the value of $R_E$ in order to change the output firing rates. Such a choice for $R_I$, used for Figure 8C-H, reflects the firing rate of the summed activity of the modeled presynaptic inhibitory neurons that is proportional, for large populations, to the (sample) mean of a neuron's presynaptic inhibitory pool. Population-averaged firing rates in rat cortex *in vivo* have been reported on the order of ~1-5 Hz, with fast-spiking putative inhibitory neurons firing faster than excitatory cells (e.g. Hengen *et al.* 2013; Buszaki & Misuzeki 2015). Within a thalamocortical projection column of the rat barrel cortex, there are ~20'000 neurons (Meyer *et al.* 2010). If 15% of those are inhibitory and their average connection probability is 1/3, the mean number of partners would be 1000, so that an average firing rate of 10 Hz per cell would yield a total inhibitory input rate of 10 kHz for the entire inhibitory pool. In Fig. 8C, we set a low excitatory rate of 1.75 kHz to explore the static f-I curve, while we varied the excitatory input rate in Figure 8D-H.

Defining $\bar{g}_{eff} = \bar{g} + \bar{g}_E + \bar{g}_I$ and $V_{eff} = \bar{g}_{eff}^{-1} \cdot (\bar{g} \cdot V_L + \bar{g}_E \cdot E_E + \bar{g}_I \cdot E_I)$ as *effective* membrane conductance and *effective* resting potential, respectively, eq. 19 can be equivalently rewritten as

(20)
$$C\frac{dV}{dt} = [\bar{g}_{eff} \cdot (V_{eff} - V) + \bar{g} \cdot \Delta \cdot e^{-(V-V_T)/\Delta}] + I_{slow} + i_{ext}$$
$$I_{slow} = \bar{g}_w \cdot w \cdot (V_w - V) \quad \tau_w \frac{dw}{dt} = (1 + e^{-(V-V_{wh})/D_w})^{-1}$$
$$i_{ext} = g_E(t)(E_E - V) + g_I(t)(E_I - V)$$

where $g_E(t)$ and $g_I(t)$ are the randomly fluctuating components of the synaptic conductances, with zero mean and with Dirac's Delta autocorrelation function.

For our numerical simulations as well as theoretical analysis of the correlation transfer in the model, we set $V = V_{eff}$ in the expression of $i_{ext}$ ignoring the multiplicative nature of the random fluctuations, which then took the approximate form

(21)
$$i_{ext} \approx [\sigma_E \cdot (E_E - V_{eff}) + \sigma_I \cdot (E_I - V_{eff})] \cdot \gamma(t)$$

Under such an approximation, the dynamical response functions $\widehat{A}(\omega)$ of the model neuron can be obtained directly, by numerical solution of the Fokker-Planck equation associated with Eqs. 20-21, as described in detail in (Ocker and Doiron, 2014, Richardson, 2009).

Similarly to the experiments, for a pair of neurons receiving partially correlated inputs, $\gamma(t)$ was chosen as

(22)
$$\gamma_1(t) = \sqrt{1-c} \cdot \eta_1(t) + \sqrt{c} \cdot \eta_c(t)$$
$$\gamma_2(t) = \sqrt{1-c} \cdot \eta_2(t) + \sqrt{c} \cdot \eta_c(t)$$

where $\eta_1(t)$, $\eta_2(t)$, and $\eta_c(t)$ are randomly fluctuating waveforms, generated as independent realizations of a Gaussian white noise, with zero mean and unitary variance, instead of an Ornstein-Uhlenbeck process to facilitate the mathematical analysis by Fokker-Planck equation (Brunel et al., 2001, Fourcaud and Brunel, 2002).

For small values of $c$, eq. 18 could then be used instead of Eq. 17 to calculate the covariance of spike counts in the two model neurons. The MATLAB scripts for reproducing Fig. 8 are available on FigShare.com.

## Results

We closely followed the approach of de la Rocha and colleagues (de la Rocha et al., 2007) and extended it by dynamic-clamp to the case of conductance inputs, in addition to conventional current inputs. We examined in the details how altering the fraction of common inputs changes the similarity of their output spike trains, in pairs of unconnected neurons.

**Cell types and electrophysiological responses.**

We recorded *in vitro* from $n = 47$ pyramidal cells and $n = 52$ interneurons in L5 of the rat primary somatosensory cortex, during the application of current- and conductance-clamp stimulation. Pyramidal cells were readily identified by the presence of a large apical trunk emanating from the cell body and oriented toward the pial surface. From an electrophysiological point of view, pyramidal neurons constituted a homogeneous group, characterized by a regular spiking phenotype with a variable degree of spike-frequency adaptation (accommodation index $0.0083 \pm 0.002$, Fig. 1A,D).

Interneurons were morphologically characterized by the absence of visible dendritic processes emanating from the cell body and presented either rounded or elongated shapes. Based on their distinct electrophysiological properties, they could be subdivided into two non-overlapping classes. The main features of the first interneuron type ($n = 23$; Fig 1B) were the short AP duration measured at half-width ($0.38 \pm 0.01$ ms), the high maximal firing rate achievable upon current injection, typically in excess of 200 Hz, the almost complete lack of spike-frequency adaptation (accommodation index $0.0014 \pm 0.0003$, Fig. 1B,E) and the high values of sag ratio ($0.94 \pm 0.005$) indicating very little sag in response to hyperpolarizing current pulses. Taken together, these features identify this neuronal type as a fast-spiking (FS) interneuron (Sippy and Yuste, 2013, Kawaguchi and Kondo, 2002, Markram et al., 2004, Petilla Interneuron Nomenclature et al., 2008, Tateno and Robinson, 2009), whose most abundant representative in the cortex is the parvalbumin-positive basket cell (Wang et al., 2002).

The second interneuron type ($n = 29$; Fig. 1C) was characterized by slightly broader APs ($0.53 \pm 0.02$ ms), sag ratios comparable to those of pyramidal cells ($0.77 \pm 0.02$), strong spike-frequency adaptation (accommodation index $0.02 \pm 0.005$, Fig. 1C,F), and rebound spikes in response to hyperpolarizing current steps, present in 20 out of 30 cells in this class (Fig. 2D). Taken together, these intrinsic properties suggest that this cell type can be identified as a non-fast-spiking (NON-FS) interneuron (Goldberg et al., 2004, Ma et al., 2006, Petilla Interneuron Nomenclature et al., 2008), such as somatostatin-positive Martinotti cells (Kawaguchi and Kondo, 2002, Silberberg and Markram, 2007, Sippy and Yuste, 2013, Wang et al., 2002).

Figures 1 and 2 summarize the main electrophysiological features of the three cell types considered in this study. As mentioned in the Materials and Methods section, for a subset of cells (13 pyramidal, 10 FS and 14 NON-FS) we measured the stationary frequency-current (f-I) curve, with and without the additional injection of background synaptic activity recreated by dynamic-clamp. FS cells had the steepest f-I curves, followed by NON-FS and by pyramidal cells (Fig. 1G-I). This ordering was observed both in the "noise-free" condition (empty markers in Fig. 1G-I)—i.e., when only DC current steps were injected into the cell—and in the condition in which the DC steps were applied on top of recreated synaptic background (solid markers in Fig. 1G-I). These results are consistent with the divisive effect on the slope of the f-I curves produced by a balanced background synaptic activity (Chance et al., 2002). Another effect attributable to the dynamic-clamp stimulation is the reduction in the measured membrane time constant. This amounted to: for pyramidal cells, $22.2 \pm 0.6$ ms to $9.8 \pm 1$ ms, $p < 10^{-10}$, for FS interneurons $6.1 \pm 0.4$ ms to $1.4 \pm 0.1$ ms, $p < 10^{-10}$ and for NON-FS interneurons $17.7 \pm 1.6$ ms to $3.9 \pm 0.3$ ms, $p < 10^{-10}$, mean ± S.E.M., two-sample Kolmogorov-Smirnov test.

**Output correlation depends on the fraction of common inputs**

We first investigated the dependence of output spike-count correlation $\rho_T$ ($T$ is the window length over which spikes are counted) on the degree of input correlation $c$, in the three cell types previously described. To this end, we performed conductance-clamp experiments and varied $c$ in the range $[0,1]$ (Fig 3A-C), while measuring spike-count correlations using a time window of length $T = 40$ ms: we chose this specific value of $T$ because it constitutes a good compromise between the detection of firing rate synchronicity (low values of $T$) and covariability (high values of $T$). In this set of experiments, we kept the firing rate of the cells constant, in the range $[10,12]$ Hz, to minimize the effects attributable to the dependency of $\rho_T$ on the cells' firing rate (de la Rocha et al., 2007) and to facilitate the comparison across cell types. The results of this first set of experiments are shown in Fig. 3: similarly to what was described previously for pyramidal cells under a current-clamp stimulation (de la Rocha et al., 2007), $\rho_T$ increases monotonically with $c$ under conductance-clamp stimulation, while always remaining smaller than the input correlation value (Fig. 3A).

Additionally, for values of input correlation lower than 0.5, $\rho_T$ depends roughly linearly on $c$ for all three cell types (Fig. 3B-D), albeit the linearity is more marked in pyramidal cells, where it extends up to $c = 0.5$. These results indicate that, in terms of input-output transfer of correlations, the three neuronal types behave similarly and that, given the linearity of $\rho_T$ for low values of input correlation $c$, the linear response theory previously developed (de la Rocha et al., 2007, Litwin-Kumar et al., 2011) and summarized in the Materials and Methods section is suited for interpreting the experimental data.

**Spike-count covariance depends on cell pairs type**

By definition, evaluating the spike-count correlation involves computing the covariance between the spike counts from neuron pairs and then normalizing by their respective variances. Linear response theory is only used to approximate the covariance (in the regime where it is small relative to the variances). By contrast, building a theory for the output variance of a neuron where complex cellular processes, such as spike frequency adaptation, that persist beyond a single spike is often difficult (Deger et al., 2014, Naud and Gerstner, 2012, Ocker and Doiron, 2014, Richardson, 2009). For this reason, we now focus on experimentally measuring and understanding the spike-count covariance between a neuron pair.

We investigated whether pyramidal cells and FS and NON-FS interneurons behave similarly in terms of covariance transfer, over a range of geometric mean firing rates. To this end, we fixed $c = 0.5$ and varied the ratio of excitatory to inhibitory presynaptic firing rates (Eq. 5) to span (geometric) mean firing rates of the pairs of cells over the range $[0,25]\ Hz$. For sufficiently irregular spike trains the absolute magnitude of spike-count covariation increases with the length of the window $T$. It is then convenient to examine the values of spike-count covariance for a given value of $T$ while comparing two conditions, e.g., different cell types or distinct stimulation paradigms.

We first found that cortical interneurons display spike-count covariance values that are higher than pyramidal cells when using a conductance clamp stimulation, over a broad range of firing rates (Fig. 4A and C) and for $T$ in the range 5 to 160 ms (Fig. 4B and D). In particular, the ratio between interneuron and pyramidal covariances is larger at higher values of geometric mean firing rates (compare diamond-shaped with asterisk markers in Fig. 4B and D), and it increases with the value of $T$, reaching values of approximately 5 and of 2 times the covariance of pyramidal cells for FS and NON-FS interneurons, respectively. Interestingly, even when considering mixed cell-type pairs (e.g., one pyramidal cell and one FS interneuron, as shown in Fig. 7A-B), the resulting covariance values are always higher than what we observed in pyramidal pairs (Fig. 7).

Taken together, these results indicate that both interneuron types have a greater capability of transmitting covariation than pyramidal cells, especially on longer timescales and for higher geometric mean firing rates.

**Spike-count covariance in pyramidal cells depends on the stimulation paradigm**

In a second set of experiments, we quantified the extent to which a conductance-clamp stimulation shapes the transfer of covariance, compared to a more conventional current-clamp stimulus. To this end, we performed additional recordings in pyramidal cells, employing the current-clamp stimulation described in the Methods, matching the protocol used by de la Rocha *et al.* (2007).

We found that for the specific window used ($T = 40$ ms), covariance transfer in conductance clamp is significantly lower than in current-clamp, at geometric mean firing rates above approximately 10 spike/s (Fig. 4E). When performing the same analysis for different values of $T$, the phenomenon of covariance shaping became particularly prominent (Fig. 4F). In particular, for short time windows $T$ the spike-count covariance was higher when employing conductance-clamp stimulation, whereas on longer timescales it was higher in current clamp, especially at higher geometric firing rates.

**Single-cell properties predict cell type-specific correlation transfer**

When input correlations $c$ are sufficiently small there is an approximate linear relation between spike correlation $\rho_T$ and $c$ (Fig. 3). Previous work has leveraged this observation to give a simple theory that expresses the spike-count covariance between two neurons $\text{Cov}(n_1^T, n_2^T)$ in terms of single-cell transfer properties (de la Rocha et al., 2007, Litwin-Kumar et al., 2011) (see Methods). For large values of $T$, the window over which the spike counts $n_1^T, n_2^T$ are estimated, this theory reduces to:

$$\text{(1)} \quad \lim_{T \to \infty} \frac{\text{Cov}(n_1^T, n_2^T)}{T} \approx c\sigma^2 \cdot \text{gain}_1 \cdot \text{gain}_2$$

where $\text{gain}_1$ and $\text{gain}_2$ are the slopes of the frequency-current curves of the two neurons. Indicating by $\mu_1$ and $\mu_2$ the mean amplitudes of the current inputs experienced by the two neurons, this is formally written as $\text{gain}_i = dr_i/d\mu_i$ for $i = \{1,2\}$. Then, the functions $n_i^T = Tr_i(I_i^T)$ link the integrated input experienced by each neuron $I_i^T = \int_0^T I_i(t)dt$ to its output spike count. Since $\mu_i = \lim_{T \to \infty} I_i^T/T$ and $r_i = \lim_{T \to \infty} n_i^T/T$ then, for large $T$, $r_i(\mu_i)$ is the frequency-current curve that we measured (see Fig. 1G-I), and $dr_i/d\mu_i$ is the firing rate gain. In other words, Eq. 1 approximates $\text{Cov}(n_1^T, n_2^T)$ as the input covariance $c\sigma^2$ scaled by the product of firing rate gains.

If $r_i(\mu_i)$ is linear, then the gain $dr_i/d\mu_i$ remains fixed as $\mu_i$ changes, so that $\text{Cov}(n_i^T, n_j^T)$ also does not shift. By contrast, a nonlinear $r_i(\mu_i)$ makes $\text{Cov}(n_i^T, n_j^T)$ depend on the "operating point" of the neuron pair (measured by $dr_i/d\mu_i$), so that $\text{Cov}(n_i^T, n_j^T)$ is malleable despite the input covariance remaining fixed at $c\sigma^2$. Intuition for this effect comes from an understanding of how the joint input density $p(I_1^T, I_2^T)$ is mapped to the joint output density $p(n_1^T, n_2^T)$. The Gaussian density of $p(I_1^T, I_2^T)$ is reshaped by the nonlinear mapping $r_1(I_1^T)$ and $r_2(I_2^T)$ of each cell in the pair, so that the density $p(n_1^T, n_2^T)$ is non-Gaussian (Fig. 8A). Nevertheless, $\text{Cov}(n_1^T, n_2^T)$ reports the effective spread of $p(n_1^T, n_2^T)$ along the diagonal $n_1 = n_2$. For low $\mu_1$ the spike threshold nonlinearity forces both the output firing rate $n_1/T$ and the gain $dr_1/d\mu_1$ to be small, compromising the spread of $p(n_1^T, n_2^T)$. When $\mu_1$ is increased, the transfer of common fluctuations is enhanced through sampling larger firing rate gains in the $n_1 = Tr(I_1^T)$ transfer. This results in an increased spread of $p(n_1^T, n_2^T)$ along $n_1 = n_2$, and naturally $\text{Cov}(n_1^T, n_2^T)$ is larger. Thus, by controlling the gain of the firing rate transfer (through $\mu_1$ and $\mu_2$) the output $Cov(n_1^T, n_2^T)$ is manipulated despite the input $\text{Cov}(I_1^T, I_2^T)$ being fixed.

We next used this theory to investigate how the single-cell properties of pyramidal, FS, and NON-FS cell types determine $\text{Cov}(n_1^T, n_2^T)$, and how it changes with firing rate. Exponential integrate-and-fire models with slow adaptation currents (Badel et al., 2008,

Brette et al., 2008) were matched to the passive and firing rate characteristics of the experimentally measured pyramidal, FS, and NON-FS cells (Fig. 8B) (see Methods for details). Mimicking the experiments, we employed a stochastic bombardment of presynaptic excitatory and inhibitory conductance inputs to drive stochastic spike train responses (Fig. 8B). The simplicity of the spike generation mechanism combined with vanishing synaptic timescales and large pre-synaptic ensembles permits a theoretical calculation of the firing rate curves $r_i(\mu_i)$ and their gains $dr_i/d\mu_i$ (see Methods). We first computed the f-I curves by injecting static currents while fixing the input variance (Fig. 4C), qualitatively matching the experimental results (compare Fig. 8C to Fig. 1G-I). The FS cells, however, had more similar f-I curves to the NON-FS and pyramidal cells in our model than in the averaged experimental f-I curves, perhaps due to the different noise levels (see Methods).

To probe the input-output transfer function, we next increased the output firing rates $r_i$ by varying the presynaptic excitatory input rate, $R_E$. For the fitted cellular and synaptic parameters the model firing rate curves qualitatively matched those reported for pyramidal, FS, and NON-FS cells. For all models the firing rate gain $dr_i/d\mu_i$ increased with the output firing rate of the cells, with FS interneurons showing the highest dependence (Fig. 8D). Furthermore, as output firing rates were driven by a stochastic bombardment of presynaptic inputs, then as the output firing rate increased so did the overall input variance $\sigma^2$, again for all models (Fig. 8E).

Finally, we modeled a pair of cells receiving correlated conductance fluctuations ($c = 0.1$) and numerically estimated $\text{Cov}(n_1^T, n_2^T)$ as firing rates varied. Our simulations recapitulated the experimental results, namely at high firing rates FS interneuron pairs showed the largest covariance, followed by NON-FS interneuron pairs, and with pyramidal cell pairs having the lowest covariance (Fig. 8F, open circles). Our theory based on single-cell properties also gave a quantitative match to $\text{Cov}(n_1^T, n_2^T)$ for all cell types (Fig. 8F, solid lines). This provides evidence that the cell-type differences in the electrical properties that determine single-cell firing rate gain (Fig. 8D) underlie, in part, the cell-type distinctions in covariance transfer (Fig. 8F).

Interestingly, the pyramidal cell model described previously qualitatively captures the shaping of the covariance values observed in the experiments (Fig. 8G-H). However, how correlation shaping depends on the cell pair firing rates is reversed in the theory compared to the experiments. More to the point, in the experiments the highest values of $Cov_T^{pyr,cc}/Cov_T^{pyr,gc}$ are observed for high values of geometric mean rate (in the range $[16,17]\ Hz$, diamond markers in Fig. 4F) and gradually decrease for decreasing geometric mean rates. In theory and in simulations, on the other hand, high geometric mean firing rates correspond to lower values of covariation shaping (diamond markers in Fig. 8H). In both theory and experiment the raw $Cov(n_1^T, n_2^T)$ values increase with firing rates (Fig 4E and 8G), yet in the theory there is a saturation in $Cov(n_1^T, n_2^T)$ with firing rates in current clamp that is absent in the *in vitro* recordings (and vice versa for the conductance-clamp case). These discrepancies may be attributed to cellular processes that are ignored in our model, such as subthreshold voltage-gate conductances whose activation differs between current- and conductance-clamp activation in real neurons.

**Measured covariance values correlate with single-cell properties**

Our linear response theory predicts that the magnitude of covariance between two spike trains from distinct neurons should depend on the slopes of their f-I curves (Eq. 23). If the f-I curves were truly linear then the spike train covariance would be a fixed value, independent of the firing rates of the neuron pair. However, in fluctuation driven regimes the f-I curves are markedly nonlinear, with a slope that grows with the firing rate (Fig. 9 A-C). Thus, by changing the firing rates of a neuron pair we sample a range of gains and associated pairwise covariances by varying the cell pair firing rates. This in turn provides an opportunity to quantitatively test our linear response prediction: for each pair of cells in our dataset we computed the coefficient of determination $r^2$ between the measured covariance values and the product of the slopes of the f-I curves of the individual cells, computed with the injection of recreated background synaptic activity, as shown in Fig. 1G-I at the population level (filled markers) and in Fig. 9A-C for the representative cells used in Fig. 9D-F. In agreement with the prediction of the linear response theory, we found a very strong correlation between the covariance and the product of the f-I curve gains, on a cell pair-by-cell pair basis (Fig. 9D-F and Fig. 10A-C), and for each of the three cell types under analysis. Interestingly, the distribution of coefficients $r^2$ (Fig. 9G-I and Fig. 10D-F) was more strongly skewed towards high values for the FS interneuron cell type, indicating that the covariation dynamics of GABAergic fast-spiking interneurons are better captured by our linear response theory, at least for the intermediate value of input correlation used in these experiments (i.e., $c = 0.5$). Taken together, these results demonstrate how microcircuit observables—the magnitude of covariation in the spike trains of two cells—can be related to intrinsic, single cell properties of the neurons that constitute the microcircuit—in this case, the slope of their f-I curves.

**Discussion**

We have shown that three distinct cell types—pyramidal cells and two types of GABAergic interneurons—in L5 of rat somatosensory cortex respond differently to the same level of correlated inputs. In our experiments, cells were stimulated with biophysically realistic conductance inputs that mimic the activation of synaptic receptors and that recapitulate the features of the high-conductance state observed *in vivo* (Destexhe et al., 2001). It is worth pointing out that our results could not have been obtained using current-clamp alone, as they crucially depend on the modulation of intrinsic cell properties that can only be ascribed to conductance injection. To the best of our knowledge, this is the first time that differences have been found in the capability of distinct neuronal types to transmit input correlations in a firing rate-dependent manner.

**Cell type-specific differences in covariation transfer**

We interpreted our experimental data through a theory that predicts how spiking neurons transfer co-fluctuations in inputs to their output spike trains (Litwin-Kumar et al., 2011). We observed that pairs of FS interneurons display a degree of covariation that is several factors larger than pairs of pyramidal cells, while pairs of NON-FS interneurons produced covariation values that were only slightly larger than pyramidal cells. Interestingly, this can be explained by an intrinsic cellular property—the greater steepness of FS interneurons' static input-output relationship than pyramidal cells—therefore tying a single-cell feature to a network observable. Although our experiments were carried out in juvenile rats, which may not have reached yet a complete degree of neuronal maturation, the relative ordering of the steepness of the f-I curves of the three cell types considered here is preserved also in adulthood (Jiang et al., 2015), thus making our main results valid beyond the juvenile stage investigated in the present study.

While our work highlights the importance of cellular differences between cell classes for correlation transfer, it ignores the specifics of how pyramidal cells and FS and NON-FS interneurons are positioned within the recurrent cortical circuit (Pfeffer et al., 2013, Tremblay et al., 2016). Linear response theory has a natural extension to recurrently coupled networks of spiking neurons (Ocker et al., 2017), where single neuron transfer functions are mixed with wiring structure to determine network correlations. However, previous studies that employ this theory have ignored differences in inhibitory neurons and rather simply model a single interneuron class (Bos et al., 2016). Furthermore, most modeling studies that include multiple interneuron subtypes have only focused on firing rate models (Del Molino et al., 2017, Kuchibhotla et al., 2016, Litwin-Kumar et al., 2016), and they do not discuss how fluctuations are distributed over a network. The cellular insights exposed in our study when combined with established recurrent circuit theory promise new insights into how biologically realistic cortical circuits produce and transfer network-wide correlations.

**Role of conductance-clamp stimulation**

Our conductance-clamp stimulation protocol was instrumental in clarifying how the modulation of spike-count covariance by the geometric mean firing rate strongly depends on the cell type. This is in contrast with what was shown in earlier experiments using a current-clamp stimulus alone (de la Rocha et al., 2007), where the authors found no major differences among regular spiking and intrinsic bursting pyramidal cells and fast-spiking interneurons.

To understand this discrepancy, we explicitly compared conductance and current-clamp stimulations in pyramidal cells (Fig. 4E-F). Our results are in agreement with the correlation-shaping mechanism first presented in (Litwin-Kumar et al., 2011), namely that high conductance states better transfers short timescale covariability, while a low conductance state (comparable to current clamp) better transfers long timescale covariability. However, in that study in vitro neurons were held at a constant firing rate, whereas here we explored a range of firing frequencies. Our work highlights that for higher geometric mean firing rates the increased covariability transfer at short timescales in the high conductance state is diminished. This should be contrasted with the increased covariability transfer at long timescales in a low conductance state (current clamp), which is enhanced at high firing rates.

In summary, our experimental and modeling results show that pyramidal neurons in the high-conductance state are more suited to transferring covariation on short timescales (i.e., via synchronous activation) than previously predicted by current-clamp experiments (de la Rocha et al., 2007). This despite the fact that the firing variability, measured as the coefficient of variation of the inter-spike intervals, is comparable in these two conditions (see Fig. 6).

Of course, the major limitation of any dynamic-clamp stimulation consists in the point nature of the conductance injection. While this might be a reasonable assumption for neurons with relatively small dendritic trees, such as basket cells (Markram et al., 2004, Wang et al., 2002), L5 pyramidal cells like the ones considered in this study have extensive dendrites containing a variety of ion channels (Harnett et al., 2015, Stuart and Sakmann, 1994), which, together with nonlinear mechanisms of synaptic integration (Harnett et al., 2012, Xu et al., 2012), support a variety of complex processing tasks (for a review see (Spruston, 2008)). This adds a further dimension to the structure of the inputs, which can be not only temporally correlated, as is the case for the inputs considered in this work, but also spatially correlated, thus leading to potentially interesting phenomena of inputs cooperativity, which largely remain to be explored.

**Relating theory and experiment**

De la Rocha and colleagues (de la Rocha et al., 2007) put forth a theory that related the gain of single neuron input-output transfer to how correlated input fluctuations are transferred by neuron pairs to output spike-count correlations (Eq. 23). However, in that study, as well as subsequent ones, *in vitro* experiments were only qualitatively compared to theory. For instance, while the dependence of spike-count correlations upon the geometric mean firing rate can be derived from our linear theory at low firing rates (Shea-Brown et al., 2008), it has only been shown to be qualitatively true in real neurons (de la Rocha et al., 2007). Similarly, theoretical work has shown how the timescale (Litwin-Kumar et al., 2011) or excitability class (Barreiro et al., 2010, Hong et al., 2012) of membrane integration determines the timescales over which correlations are transferred, nevertheless experimental tests were only qualitative in nature (e.g., an increase or decrease in correlation as membrane properties are changed).

The difficulty with a quantitative test of a linear response theory of correlations is that a systematic measurement of both neuronal gain and correlations transfer in neuron pairs is rarely performed over a range of firing rates. Indeed, studies that focus upon single neuron gain control explicitly ignore populations of neurons, while studies of neuronal populations often only study the network in a single regime (and hence a single gain value). The results presented in Fig. 9 of our study thus constitute the most experimentally validated test of this theory, comparing the mapping between neuronal gain and covariability over a range of gain values. Furthermore, our work uses the heterogeneity over pyramidal cells and FS and NON-FS interneurons to test our theory over a broad range of cellular properties.

**Functional implications**

Overall, our results raise the intriguing possibility that principal cells in the neocortex—and possibly in other areas of the brain, such as the hippocampus—might be specifically tuned to operate in a low-correlation level, which could improve the coding capabilities of a population of principal cells. On the other hand, the higher covariance values measured in the two interneuron classes, and in particular in fast-spiking interneurons, are likely to play a role in the regulation of the activity of local circuits by interneurons. In particular, higher level of correlation might facilitate interneurons in their function as providers of a "blanket of inhibition" (Karnani et al., 2014, Packer and Yuste, 2011), and, in the case of FS interneurons, could be implicated in the important role these cells play in orchestrating cortical oscillations (see (Freund and Katona, 2007) and references therein). Interestingly, it has recently been shown that a class of GABAergic cells in the prefrontal cortex sends long-range projections to subcortical areas (Lee et al., 2014): it would be of great interest to investigate whether these cells present correlation transfer properties that differ from what we have described here.

In conclusion, our findings underline the importance of an in-depth characterization of cortical cell diversity by increased experimental realism and point to a richness of network behaviors arising from the diversity of intrinsic cell properties.

## Contributions

DL, BD, and MG designed and supervised the research. DL performed and analyzed the electrophysiological experiments, GKO performed numerical simulations and contributed the theoretical predictions. DL, GKO, BD, and MG wrote the paper. All authors read and approved the final manuscript.

## Data accessibility statement

Authors confirm that the data underlying their findings are fully available. Relevant data sets and analysis scripts have been stored at FigShare.com (DOI m9.figshare.7241144), including an index of the deposited data.

**Figure**

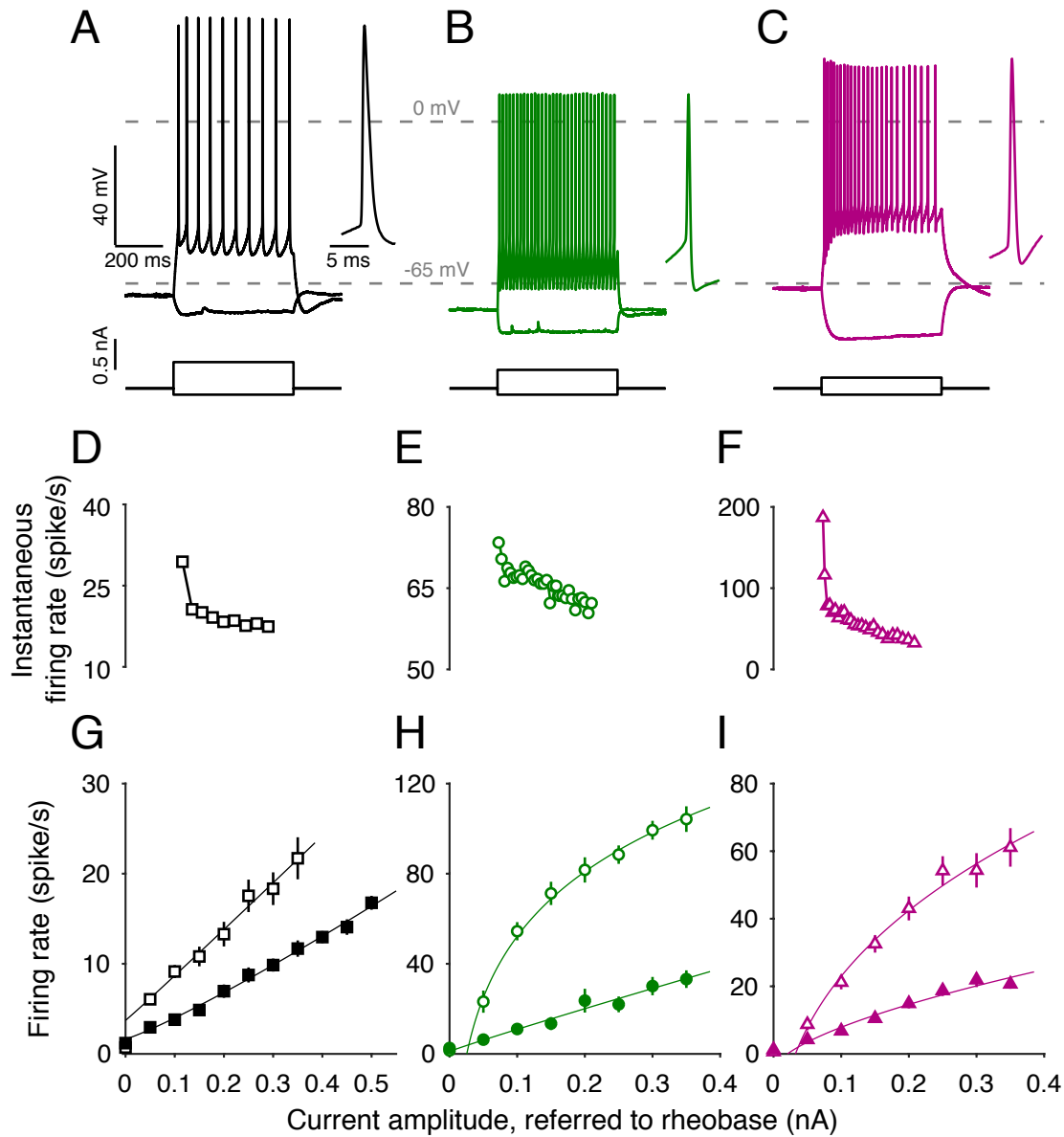

**Figure 1. Electrophysiological intrinsic properties of pyramidal cells and two types of interneurons.** (A) Typical membrane potential of pyramidal cells in response to the injection of hyperpolarizing and depolarizing current steps. Inset: magnification of the first action potential fired by the cells. (B) Same as (A) but for fast-spiking interneurons with non-accommodating firing pattern and (C) for low-threshold-spiking, non-fast-spiking interneurons with accommodating firing pattern. (D) Instantaneous firing rate, computed as the inverse of the inter-spike intervals, in response to the depolarizing current steps

displayed in A. (E-F) Same as D, but for FS and NON-FS interneurons, respectively. (G) Population summary of the stationary transfer function for pyramidal cells (N = 13) in response to the injection of depolarizing current steps, with and without the background synaptic activity recreated in dynamic-clamp (solid and empty markers, respectively). The solid lines are power-law fits of the form $f(I) = a \cdot I^b + c$, with *a, b,* and *c* free parameters. Note both the divisive effect of the recreated synaptic background on the slope of the current-frequency curves, as well as the overall diversity of the average slope of the curves across cell types. (H-I) Same as G, but for FS (N = 10) and NON-FS (N = 14) interneurons, respectively. Figure 2 presents the distinct electrophysiological features of the three cell types.

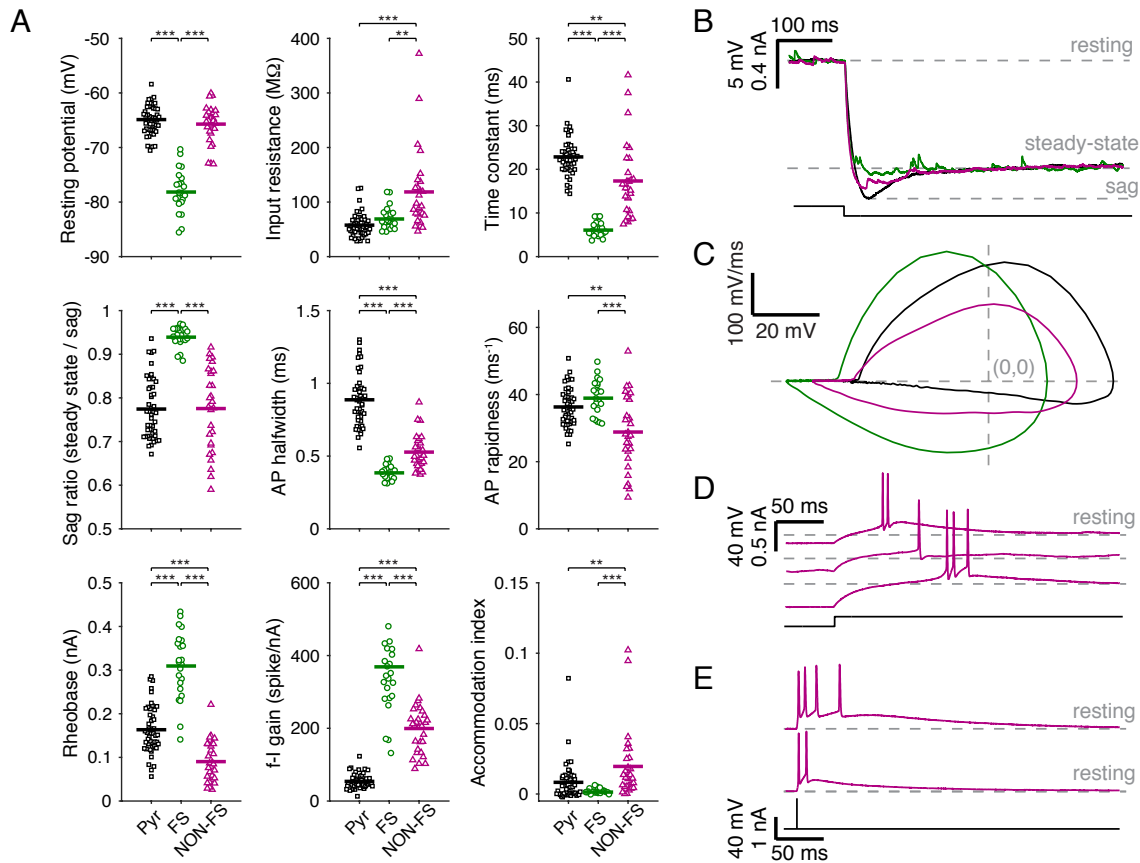

**Figure 2. Distinct electrophysiological features of three cortical cell types.** Our classification of pyramidal cells, FS and NON-FS interneurons is supported by a well-known heterogeneity of passive and active electrophysiological observables, as reported in the literature for basket cells and Martinotti cells and shown in (A) for our dataset. These observables were compared across 47 pyramidal cells, and 23 FS and 29 NON-FS interneurons, and their statistically significant differences complement the diversity of transient and stationary firing transfer properties described in Fig. 1. Two and three asterisks indicate a value of $p < 0.01$ and $p < 0.001$, respectively, for statistically significant differences as assessed by a Kruskall-Wallis test. (B) Representative voltage deflections in the three cell types in response to a step of current that caused a hyperpolarization of approximately 10 mV. Note the comparable amount of sag in the response of the pyramidal cell and NON-FS interneuron, and the almost complete lack of sag in the response of the FS interneuron. The membrane voltages have been vertically aligned to the value of the resting potential before the stimulus onset, for ease of comparison. (C) Representative voltage traces in the plane $dV/dt$ *versus* $V$ of the average action potential for the three cell types. Note how pyramidal cells display the largest action potentials, with onset speeds comparable to those of FS interneurons, which however are also characterized by an extremely fast AP offset. (D-E) Rebound firing following hyperpolarization and burst generation in response to brief current pulses have been reported as distinctive features of somatostatin-positive NON-FS interneurons: these properties were observed in our experiments for NON-FS cells only, as shown here in 5 representative cells.

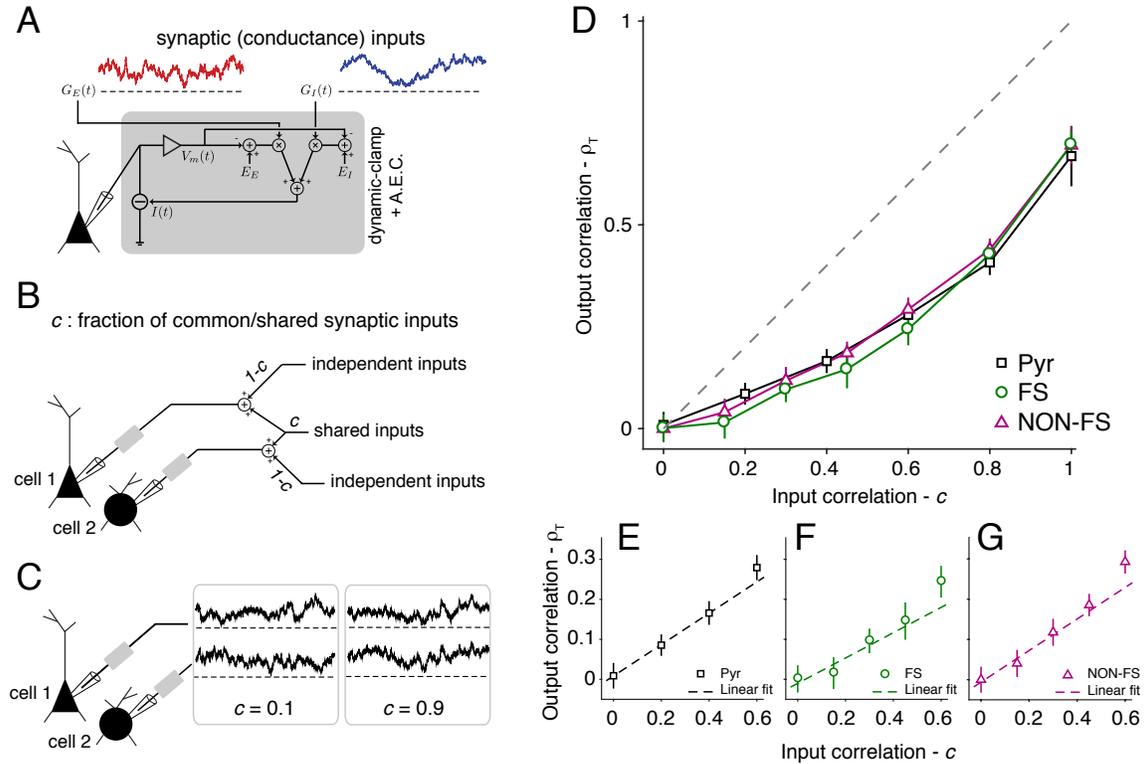

**Figure 3. Output correlation $\rho_T$ as a function of input correlation $c$, for three cortical cell types.** Two sets of conductance-based fluctuating waveforms, sharing a common fraction $c$, were applied to pairs of anatomically unconnected neurons. At the steady-state, the correlation coefficient $\rho_T$ between their spike responses was estimated over a spike-counting window of $T = 40$ ms and found to increase monotonically with increasing input correlation $c$, as intuitively expected. (A) Schematic of the dynamic-clamp experimental setup: the total current injected into the cell is the sum of two conductance waveforms, one excitatory (red trace) and one inhibitory (blue trace), each multiplied by the appropriate driving force. This in turn depends on the instantaneous membrane potential, recorded in real-time by the system and filtered by the Active Electrode Compensation. (B) The stimulation to each cell in an unconnected pair comprises an independent and a shared component. The ratio of these two is regulated by the correlation coefficient $c$. (C) Effect of varying $c$ on the conductance waveforms injected into the two cells. Notice the high (low) degree of similarity of the traces corresponding to $c = 0.9$ ($c = 0.1$). (D) Mean values of $\rho_T$ for $T = 40$ ms over the possible range of $c$ for pyramidal cells (black markers), FS (green markers) and NON-FS interneurons (purple markers) cells. Error bars indicate

the standard deviation. The dashed grey line has a unitary slope and highlights how the output correlation is always smaller than the input correlation. (E) For weak input correlations ($c \leq 0.5$), the output correlation in pyramidal cells displayed a direct proportionality to $c$, a useful property for the subsequent theoretical interpretation of our results. Shown is the same data as in (D), magnified in the interval $c \in [0, 0.6]$ and fit with a linear function of the form $\rho_T(c) = m\,c$, where $m$ is the proportionality coefficient. The fit was performed in the interval $c \in [0, 0.4]$ to highlight how the value of $\rho_T$ corresponding to $c = 0.6$ departs from the linear relationship between input and output correlation. The data is the average of $N = 6$ cells firing at a firing rate $\nu = 12.5 \pm 2.2$ spike/s. (F) Same as (E), but for FS interneurons ($N = 5$ and $\nu = 11 \pm 2.8$ spike/s). (G) Same as (E), but for NON-FS interneurons ($N = 6$ and $\nu = 10 \pm 1.7$ spike/s).

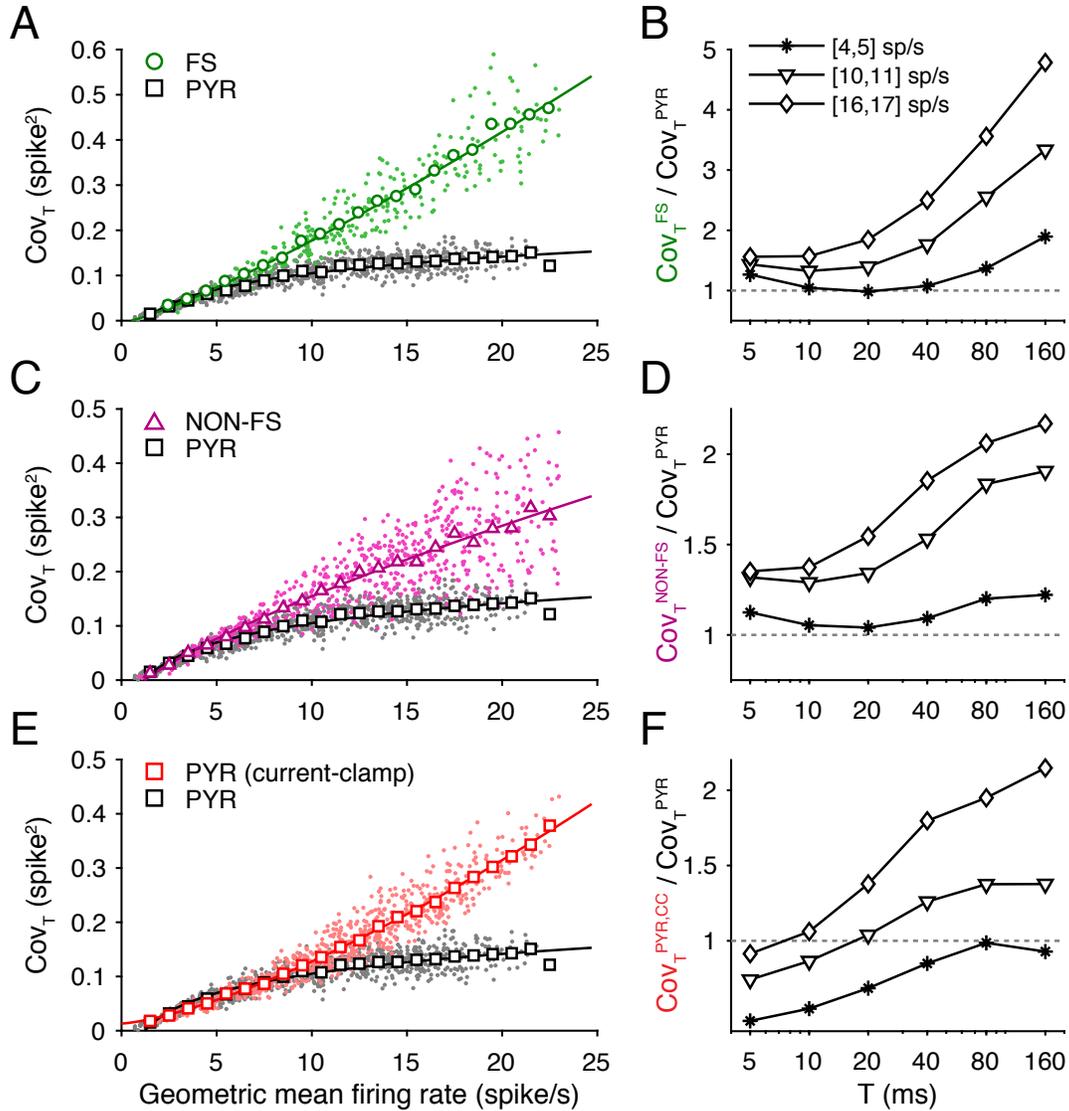

**Figure 4. Diversity of correlation transfer across cell type.** The covariance of the output spike count was estimated in cell pairs of identical type, for a fixed input correlation $c = 0.5$, across distinct firing regimes (A,C,E) and spike-count window sizes (B,D,F). (A) Each dot is a value of covariance for a given pair of cells and repeated stimulation block, computed over a count window $T = 40$ ms and for increasing mean firing rates. Large markers represent the mean of all the individual values, taken in 1 spike/s bins, revealing over our entire data set a cell-type specific correlation transfer. The solid lines are optimal fits of the binned data with a power-law function of the form $Cov(f) = a \cdot f^b + c$, where $f$ is the geometric mean firing rate of the pair and $a$, $b$ and $c$ are free parameters. Grey dots

and black markers and lines refer to pyramidal cells (N=25, with a total number of 855 pairs), while light green dots and green markers and lines refer to FS interneurons (N=14, with 285 total number of pairs). (B) Quantification of covariance shaping in pyramidal cells and FS interneurons, across different spike counting window sizes, for firing frequency regimes in the ranges [4,5], [10,11] and [16,17] spike/s: note how FS interneurons transfer up to 4 times more input correlations than pyramidal cells, over long time-scales and for similar firing rates. (C-D) Same as (A-B), but comparing pyramidal cells and NON-FS interneurons, shown with light purple dots and purple markers and lines (N=15, with a total number of 621 pairs). (E-F) Same as (A-B), but comparing pyramidal cells subjected to conductance-clamp (same data as in panels A and B) and current clamp stimulation, shown with light red dots and red markers and lines (N=11, with a total number of 716 pairs). Figures 5 and 6 report representative membrane potential traces, the coefficient of variation of the interspike interval distributions, respectively. Figure 7 repeats the analysis for mixed cell-pairs.

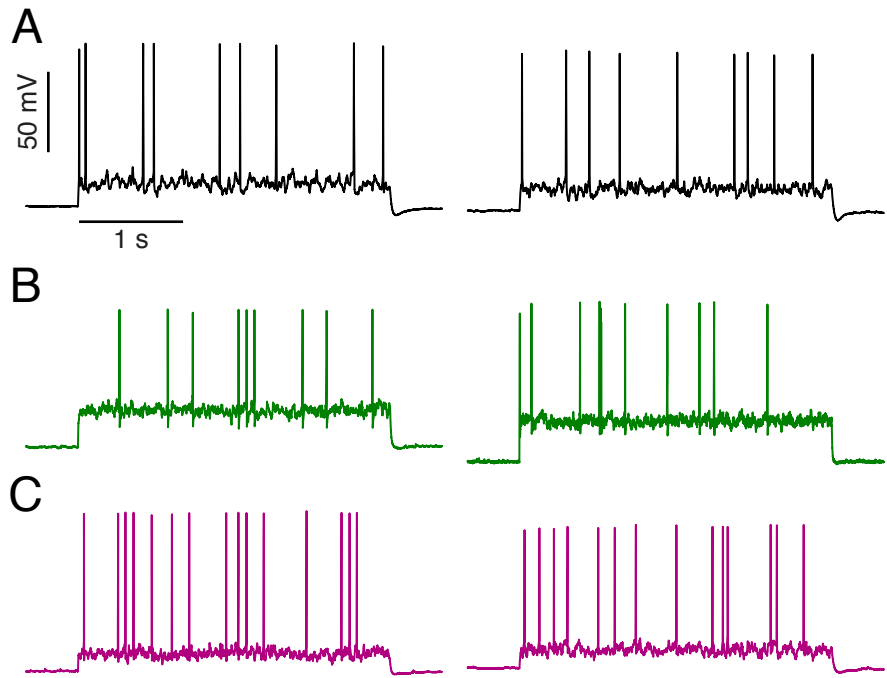

**Figure 5. Representative voltage traces in response to recreated synaptic stimulation in the three cell types.** The average barrage of recreated excitatory and inhibitory synaptic

inputs was balanced at a voltage that elicited a firing rate in the range [3,5] spike/s. (A) Responses of two pyramidal neurons. (B) Responses of two FS cells. Note how the noisy stimulation employed here allows obtaining arbitrarily low firing rates, effectively turning this cell type into a type-1 oscillator. (C) Responses of two NON-FS interneurons. Note how the discharge patterns more closely resemble those of the pyramidal cells.

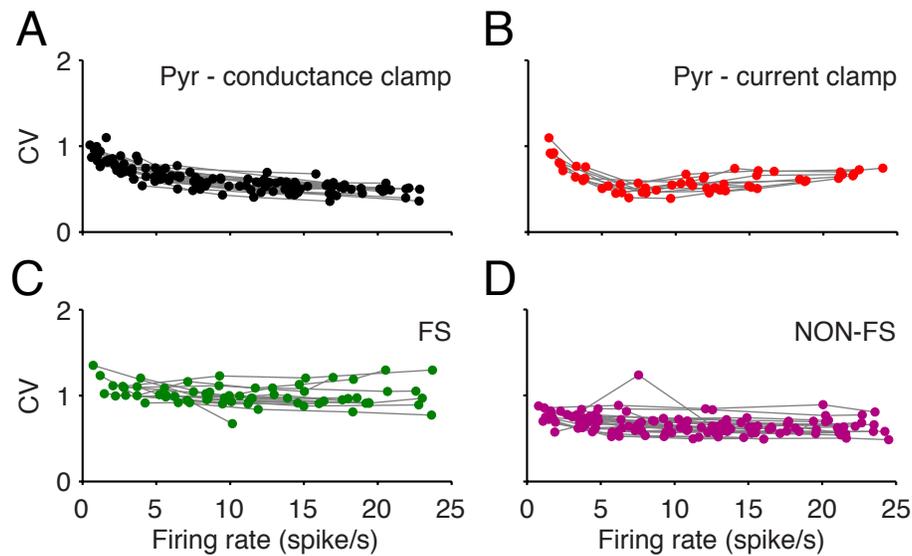

**Figure 6. Coefficient of variation (CV) of the inter-spike intervals (ISIs) across cell types, upon conductance and current stimulation.** For each stimulation block used to compute the values of covariance shown in Fig. 4, we extracted the average CV of the ISIs and plotted it against the mean firing rate of the cell in that specific stimulation block. (A) CV as a function of mean firing rate for pyramidal cells upon conductance clamp stimulation. (B) CV as a function of mean firing rate for pyramidal cells upon current clamp stimulation. (C) Same as (A), but for FS interneurons. (D) Same as (A), but for low-threshold-spiking NON-FS interneurons.

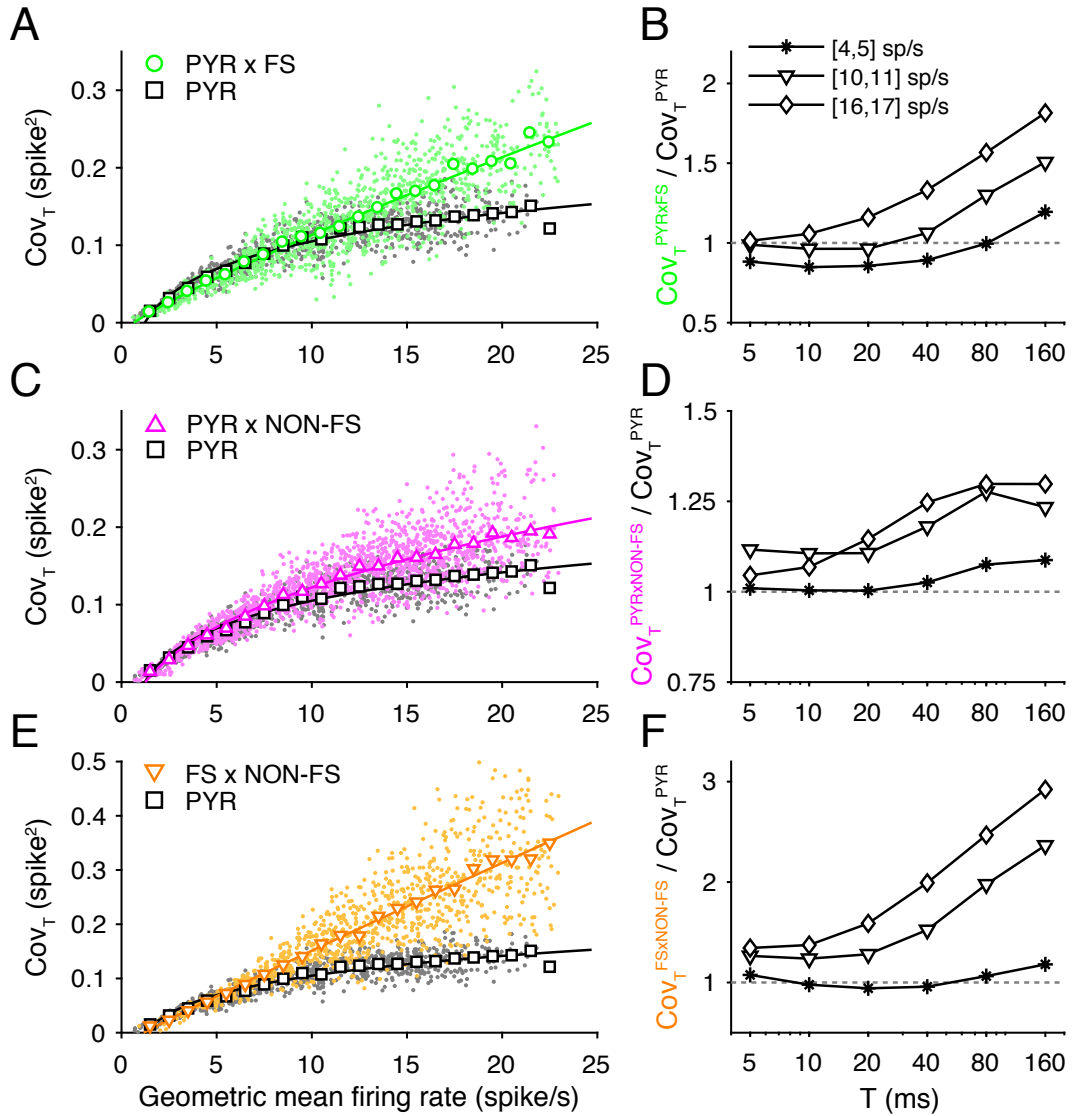

**Figure 7. Correlation transfer in mixed cell-pairs.** (A,C,E) Covariance values computed in heterogeneous pairs of cells: pyramidal and FS interneurons (A, green markers and traces), pyramidal and NON-FS interneurons (C, purple markers and traces) and FS and NON-FS interneurons (E, orange markers and traces). Notice how the resulting covariance is always greater than that computed in homogenous pairs composed of two pyramidal cells (black markers and traces in all panels, same data as in Fig. 4). (B,D,F) Covariance shaping of mixed pairs of cells. Same analysis as that of Fig. 4.

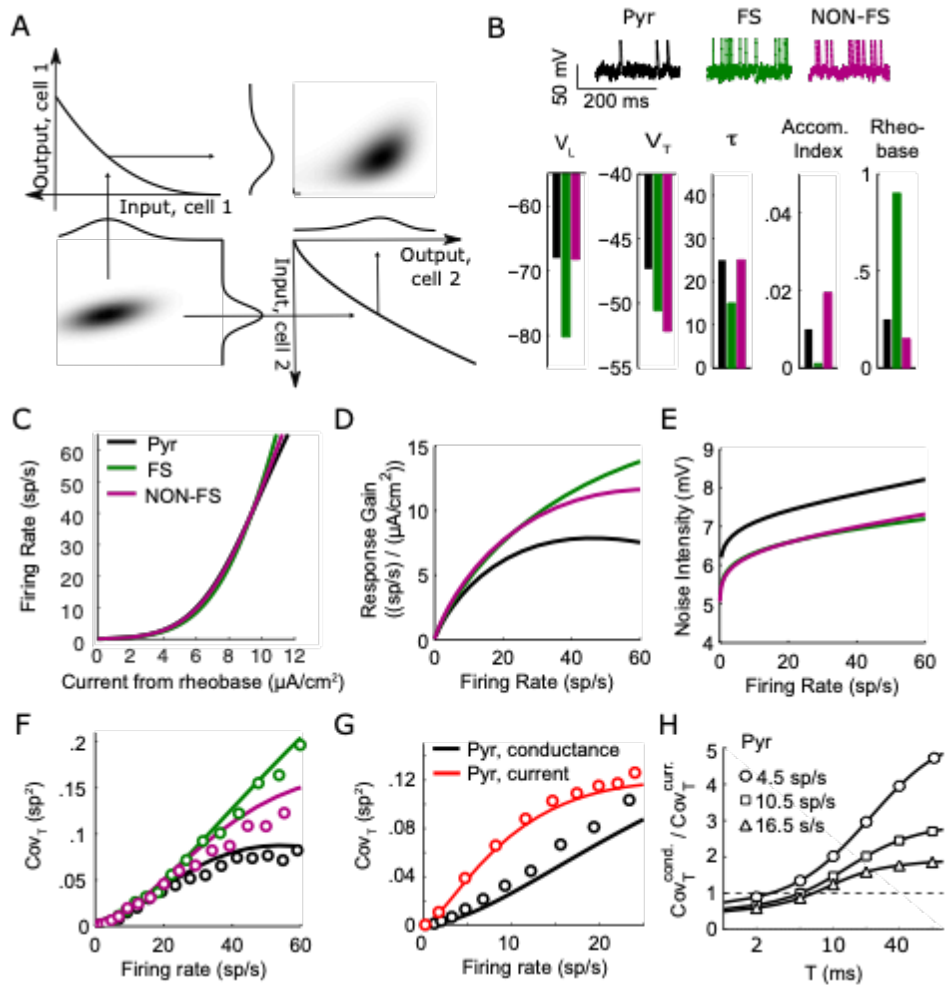

**Figure8: Single-cell properties predict covariance transfer.** (A) Schematics of covariance transfer with two cells. A correlated Gaussian input (bottom left) is transferred through each cell's f-I curve (top left, bottom right) to produce a correlated, non-Gaussian output (top right). Linear response theory predicts the covariance of that output distribution by linearizing the f-I curves around the stimulus mean. (B) Cell type-specific adapting exponential integrate-and-fire models match the recorded single-cell properties. Units: Leak reversal potential ($V_L$) and spike threshold ($V_T$) in mV; passive membrane time constant ($\tau$) in ms; rheobase in nA. (C) Static transfer functions (f(I) curves) for each model. (D) Slope of the f-I curve as a function of firing rate when the rate of the presynaptic excitatory population is increased (Methods: Neuron models). (E) Intensity of the input noise as the rate of the presynaptic excitatory population is increased. (F) Spike count covariance *versus* geometric mean output rate for each cell type. (G) Spike count covariance *versus* firing rate.

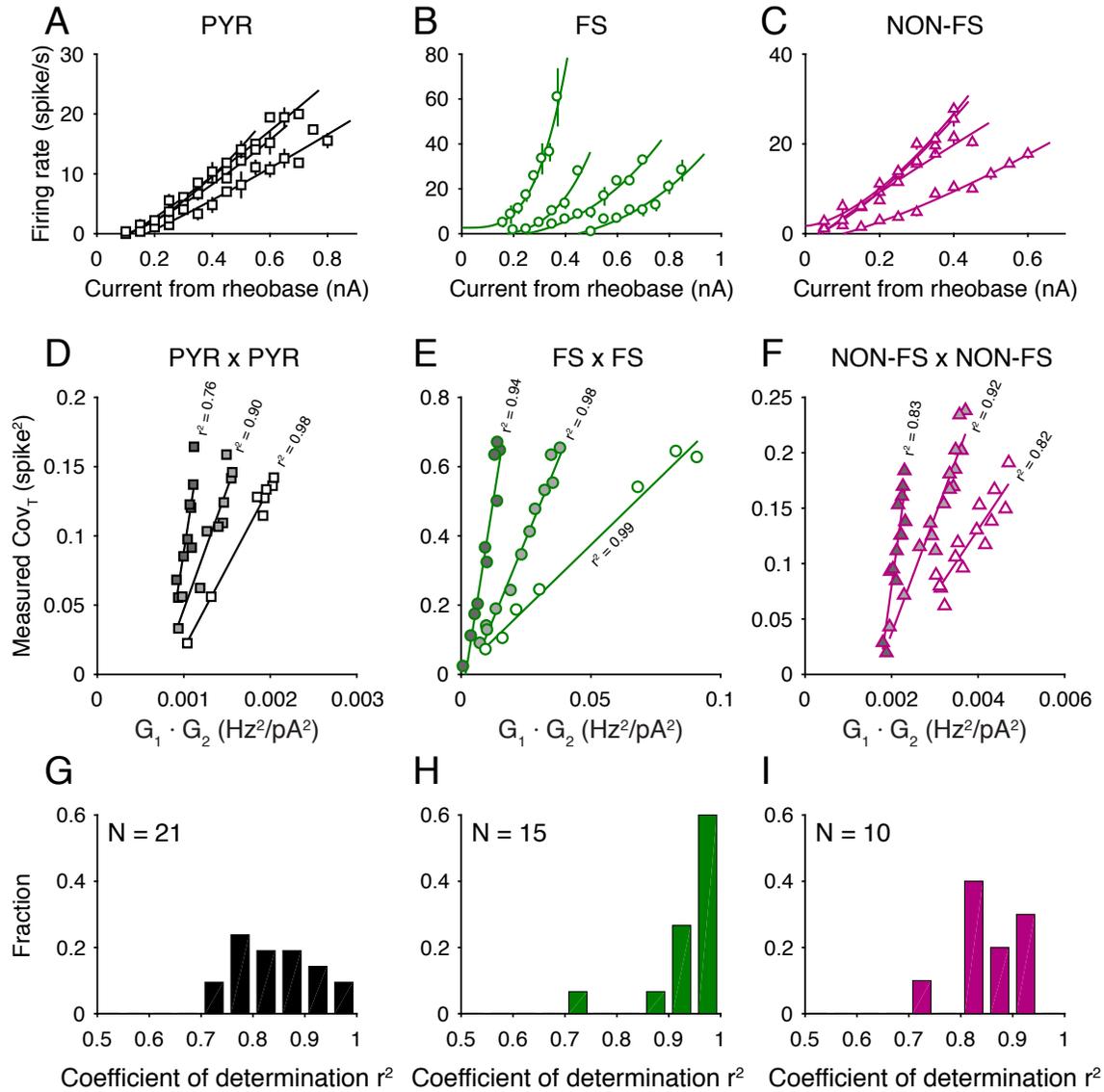

**Figure 9. The product of the slopes of the f-I curves strongly correlates with the measured covariance values, on a cell-pair by cell-pair basis and in a frequency-dependent manner.** For each cell pair, we computed several covariance values corresponding to different geometric mean firing rates. We plotted them as a function of the product of the slopes of the f-I curves of each cell of the pair (i.e. $gain_1$ and $gain_2$), at the corresponding value of firing rate. This uncovers a strong correlation between measured

covariance and product of gain$_1$ and gain$_2$, as predicted by the linear response theory. (A) Individual f-I curves upon injection of constant steps of current in addition to recreated background synaptic activity for the pyramidal cells shown in D. (B-C) Same as A, but for the FS and NON-FS cells whose covariance values vs. product of f-I gains are shown in E and F, respectively. (D) Representative examples of correlation between measured covariance and slopes of the f-I curves, for three pairs of pyramidal cells. Actual values are shown with square markers, while solid lines are linear fits to the data. The r$^2$ coefficients are indicated in the panel for each cell pair. (E-F) Same as D, but for FS and NON-FS interneurons, respectively. (G) Distribution of r$^2$ coefficients for $N = 20$ pairs obtained from 8 pyramidal cells. (H-I) Same as G, but for FS ($N = 15$ pairs from 6 cells) and NON-FS interneurons ($N = 10$ pairs from 5 cells), respectively. The value of the spike counting window is $T = 40$ ms in all panels. Figure 10 repeats this analysis for mixed cell-pairs.

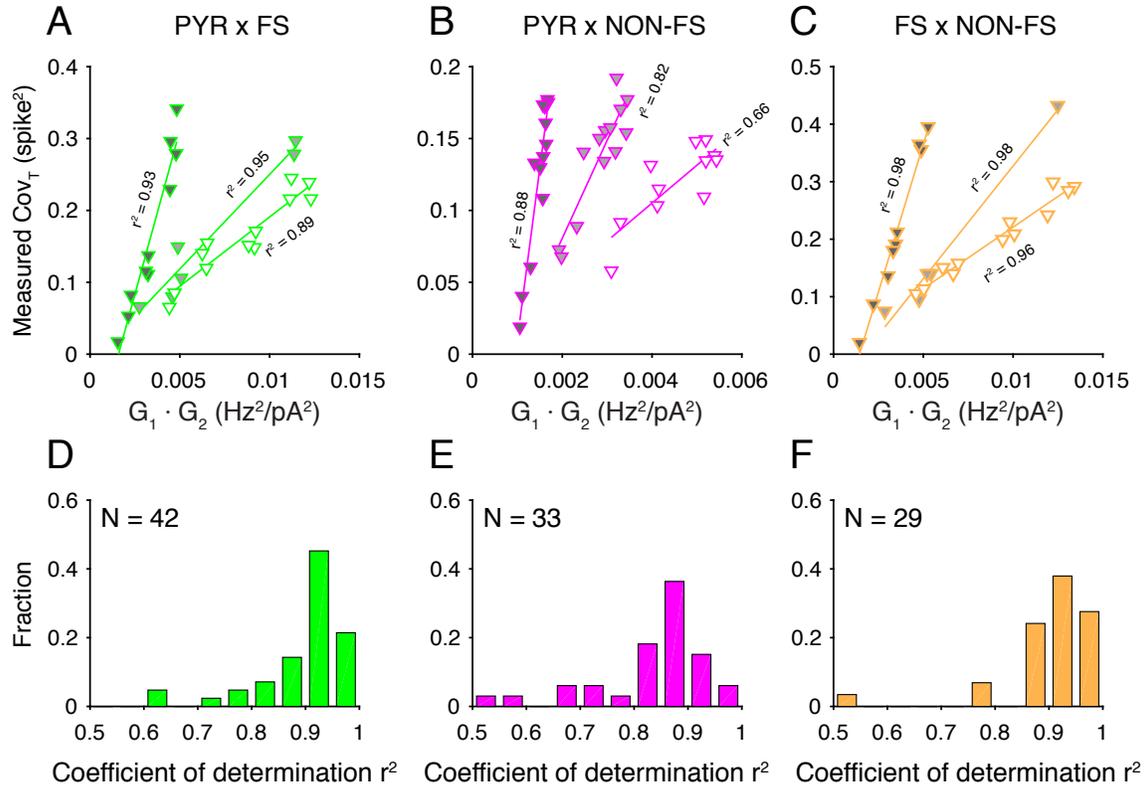

**Figure 10. Correlation between the product of the gain of the f-I curves and the measured covariance values, in mixed cell pairs.** Same analysis as that performed in Fig. 9, but for pairs of cells composed of different cell types. Also in this case there is a very strong correlation between the experimentally measured values of covariance and the product of the slopes of the f-I curves of the two cells in the pair.